\documentclass[11pt,aps,nofootinbib,prd,aps,epsf,floats,axodraw,
               amsmath,amssymb,amsfonts]{revtex4}
\usepackage{amsmath, amssymb}
\newcommand{\mathsym}[1]{{}}

\usepackage{graphicx}
\usepackage{amsmath}
\usepackage{amssymb}
\usepackage{bm}
\newcommand{\be}{\begin{equation}}
\newcommand{\ee}{\end{equation}}
\newcommand{\bea}{\begin{eqnarray}}
\newcommand{\eea}{\end{eqnarray}}

\newcommand{\rem}[1]{}
\newsavebox{\PSLASH}
 \sbox{\PSLASH}{$p$\hspace{-1.8mm}/}
 
\renewcommand{\theequation}{\thesection.\arabic{equation}}
\newcounter{saveeqn}
\newcommand{\add}{\addtocounter{equation}{1}}
\newcommand{\alpheqn}{\setcounter{saveeqn}{\value{equation}}%
\setcounter{equation}{0}%
\renewcommand{\theequation}{\mbox{\thesection.\arabic{saveeqn}{\alph{equation}}}}}
\newcommand{\reseteqn}{\setcounter{equation}{\value{saveeqn}}%
\renewcommand{\theequation}{\thesection.\arabic{equation}}}

 \newsavebox{\notrightarrow}
 \sbox{\notrightarrow}{$\to$\hspace{-4mm}/}
 
 \newsavebox{\PARTIALSLASH}
 \sbox{\PARTIALSLASH}{$\partial$\hspace{-1.6mm}/}
 
 \newsavebox{\ASLASH}
 \sbox{\ASLASH}{$A$\hspace{-2.1mm}/}
 
 \newsavebox{\KSLASH}
 \sbox{\KSLASH}{$k$\hspace{-1.8mm}/}
 
 \newsavebox{\LSLASH}
 \sbox{\LSLASH}{$\ell$\hspace{-1.8mm}/}
 
 \newsavebox{\QSLASH}
 \sbox{\QSLASH}{$q$\hspace{-1.8mm}/}
 
 \newsavebox{\DSLASH}
 \sbox{\DSLASH}{$D$\hspace{-2.2mm}/}
 
 \newsavebox{\DbfSLASH}
 \sbox{\DbfSLASH}{${\mathbf D}$\hspace{-2.8mm}/}
 
 \newsavebox{\DELVECRIGHT}
 \sbox{\DELVECRIGHT}{$\stackrel{\rightarrow}{\partial}$}
 
 \newcommand{\blue}{\IfColor{\textCadetBlue}{}}
\newcommand{\black}{\IfColor{\textBlack}{}}
\newcommand{\red}{\IfColor{\textRed}{}}
\newcommand{\green}{\IfColor{\textOliveGreen}{}}
\newcommand{\lila}{\IfColor{\textRedViolet}{}}

\begin{document}
\begin{flushright}
 [math-ph]
\end{flushright}
\title{$\alpha^*$-Cohomology,
and
Classification of
\par Translation-Invariant Non-Commutative Quantum Field Theories
}

\author{Amir Abbass Varshovi}\email{amirabbassv@ipm.ir}

\affiliation{
   School of Mathematics, Institute for Research in Fundamental Sciences (IPM).\\
   School of Physics, Institute for Research in Fundamental Sciences (IPM).\\
                                     Tehran-IRAN}
\begin{abstract}
       \textbf{Abstract\textbf{:}} Translation-invariant $\star$ products are studied in the setting of $\alpha^*$-cohomology. It is explicitly shown that all quantum behaviors including the Green's functions and the scattering matrix of translation-invariant non-commutative quantum field theories are thoroughly characterized by $\alpha^*$-cohomology classes of the star products.
\end{abstract}
 \maketitle

\section{Introduction}\label{introduction}
 \indent  Non-commutative quantum field theory is the most generic shared domain of quantum physics and non-commutative geometry. There are also various fascinating applications of non-commutative geometry in quantum physics such as to describe the Standard Model in the setting of non-commutative Riemannian manifolds or spectral triples \cite{CM}, but the naive idea of non-commutative space-time was strongly verified by other understandings of quantum physics. More precisely, the proposal of non-commutative structures for space-time coordinates at very small length scales, was in fact suggested long time before appearing the ideas of non-commutative geometry, in the early years of quantum mechanics by its founding fathers to introduce an effective ultraviolet cutoff for quantum field theories and to give an appropriate setting to describe the small scale structures of the universe \cite{Schrodinger, Heisenberg}. However, the appearance of non-commutative geometry in quantum physics led to a revival of this idea for non-commutative space-time at Planck scale which was largely ignored in the mid of the last century due to the success of renormalization program of field theories \cite{CDS}. But in fact this revival also owes most of its appearance to developments of string theory where more evidences for non-commutative space-time came from \cite{Witten, Veneziano, GM, ACV}. Indeed, in string theory as an appropriate modification of classical general relativity, the need of non-commutative space-time is actually more apparent than in quantum field theory \cite{SW}. In fact the occurrence of non-commutative field theories can also be explained in this setting by open string degrees of freedom known as D-branes \cite{Polchinski}, which are fixed hyper-surfaces in space-time onto which the end-points of strings can attach. Actually the low-energy effective field theory of D-branes has configuration space which is described in non-commuting space-time coordinate fields \cite{Witten2, Sheikh-Jabbari}.\\
  \par Algebraic formulation of non-commutative space-time coordinates was firstly worked out by Snyder \cite{Snyder1, Snyder2} which was strongly motivated by the need to control the divergences of quantum electrodynamics in its very beginning formalisms. Soon after by following the idea of Weil-Wigner quantization \cite{Weyl, Wigner} a more spectacular framework of non-commutative structure for space-time, so called Groenewold-Moyal product, was introduced \cite{Groenewold, Moyal} which after 20 years led both to a deep understanding of quantum physics and a fundamental revolution in Poisson geometry and Poisson algebra known as the theory of deformation quantization or star products \cite{BFFLS1, BFFLS2}. Non-commutative field theories were extensively studied in this generic approach, using the well-known Groenewold-Moyal product instead of the ordinary product. Developing the Feynman rules \cite{Filk} and considering the perturbation theory \cite{MVS}, revealed a serious problem in renormalization program of non-commutative scalar field theories, so called UV/IR mixing. UV/IR mixing is a pathological effect which plagues the theory by reflecting UV divergences in new IR singularities. Soon after it was shown \cite{Hayakawa1, Hayakawa2, GKW, MST} that non-commutative gauge theories are also suffered by UV/IR mixing. Curing UV/IR mixing in non-commutative quantum field theories led to more serious problems for breaking the translation-invariance \cite{GW1, GW2} or the locality \cite{GMRT}. Breaking the translation-invariance plagues a quantum theory by losing the energy-momentum conservation law. This fact provided strong motivations for studying a large family of generalized Groenewold-Moyal star product which do not depend on coordinate functions. This family of star products is commonly referred to as translation-invariant products \cite{GLV, Galluccio} since they keep the translation-invariance property of the theory when they are used instead of the ordinary product. Translation-invariant quantum field theories equipped with translation-invariant non-commutative products are also called translation-invariant non-commutative quantum field theories. It is well known \cite{GLV} that at one-loop corrections the quantum behaviors of translation-invariant non-commutative scalar and gauge field theories are entirely described by the commutators of coordinate functions. This fact can be considered as a special case of $\alpha$-cohomology description of translation-invariant star products introduced in \cite{GLV}. By definition $\alpha$-cohomology is a theory of cohomology based on the associativity condition of algebraic products which classifies the set of translation-invariant products up to a family of commutative ones, the coboundaries \cite{Varshovi}. Using the perturbation theory, it is seen \cite{TV} that the quantum corrections of translation-invariant non-commutative scalar field theory for 2- and 4-point functions, in all finite order of loop calculations can be described by the $\alpha$-cohomology classes of the $\star$ products.\\
  \par In this paper, translation-invariant star products are discussed in the setting of $\alpha$- and $\alpha^*$-cohomology in order to obtain a consistent framework to classify and study the translation-invariant non-commutative field theories. This leads to a deep understanding of quantization of translation-invariant field theories and their intrinsic non-commutative effects such as non-locality and UV/IR mixing.\\
  \par The article is arranged as follows: The second section is a short survey on translation-invariant star products due to \cite{GLV, Galluccio}. In the third section $\alpha$-cohomology is defined in a more strict and well-defined mathematical setting. It is then proven that any commutative translation-invariant star product is exactly a coboundary element and consequently it is inferred that the second $\alpha$-cohomology group strictly classifies non-commutative translation-invariant star products modulo the commutative ones, the fact of which was not shown in \cite{GLV}. In the fourth section an algebraic version of Hodge theorem is worked out for the second $\alpha$-cohomology group. This leads to a unique representing element, so called harmonic form, for each $\alpha$-cohomology class. Considering complex products results in a new and more effective cohomology theory, the $\alpha^*$-cohomology, as a sub-theory of $\alpha$-cohomology, which is introduced and discussed in section $\emph{\emph{V}}$. Using the fruit-full concept of harmonic forms for $\alpha^*$-cohomology, it is shown in the sixth section that the quantum corrections of translation-invariant non-commutative quantum field theories are entirely described by the $\alpha^*$-cohomology classes in all orders of loop calculations. This produces a non-perturbative proof and a more compact explanation for the correlation of $\alpha^*$-cohomology second group and the classification of quantum corrections in translation-invariant non-commutative quantum field theories which was partly discussed in \cite{TV}. In the seventh section the origin of $\alpha$- and $\alpha^*$-cohomology is discovered. This power-full achievement naturally shows that classifying the translation-invariant non-commutative quantum field theories in the setting of $\alpha^*$-cohomology theory is the strongest and also the most general classification in the viewpoint of quantum physics. Eventually a few number of use-full algebraic structures of translation-invariant products are worked out in section $\emph{\emph{VIII}}$. Section $\emph{\emph{IX}}$ includes the summery and conclusions and finally the last section devotes to appendices.\\


  \par
  \section{Translation-Invariant $\star$ Products}
  \setcounter{equation}{0}
  \par As a generic definition a non-commutative structure on $\mathbb{R}^{2n}$, is usually given by a set of nontrivial commutation relations of coordinate functions of a fixed globally defined chart, say $(\mathbb{R}^{2n},\{x^\mu\}_{\mu=1}^{2n})$, with an anti-symmetric fixed matrix $\theta$;
\begin{eqnarray}\label {1}
[x^\mu,x^\nu]=i\theta^{\mu\nu}~,
\end{eqnarray}
\par \noindent $1\leq \mu,\nu\leq2n$. It is seen that commutation relations (\ref {1}) can be satisfied by replacing the ordinary product of $C^\infty (\mathbb{R}^{2n})$ by a non-commutative one, say $\star$;
\begin{eqnarray} \label {2}
x^\mu \star x^\nu -x^\nu \star x^\mu=i\theta ^{\mu\nu}~.
\end{eqnarray}
 \par Usually the deformation quantization structure $\star$ is considered as the well-known Groenewold-Moyal product; i.e.; $f\star_{G-M} g=\mathfrak{m} \circ (\exp(\frac{1}{2}i\theta^{\mu\nu} \partial_\mu \bigotimes \partial_\nu) (f \bigotimes g))$ for $f,g\in C^\infty (\mathbb{R}^{2n})$ and for $i\theta$ a fixed Hermitian matrix proportional to $i\left(
        \begin{array}{cc}
          0 & 1 \\
          -1 & 0 \\
        \end{array}
      \right)
 $ for any 2-dimensional non-commutative subspace while $\mathfrak{m}:C^\infty (\mathbb{R}^{2n})\bigotimes C^\infty (\mathbb{R}^{2n})\rightarrow C^\infty (\mathbb{R}^{2n})$ is the ordinary point-wise production. The simplest generalization of Groenewold-Moyal product is the Wick-Voros production \cite{Bayen, Voros, BW1, BW2} which is defined similar to the Groenewold-Moyal one but with replacing  $i\left(
        \begin{array}{cc}
          0 & 1 \\
          -1 & 0 \\
        \end{array}
      \right)$ by $\left(
                     \begin{array}{cc}
                       1 & i \\
                       -i & 1 \\
                     \end{array}
                   \right)
      $. It can also be shown that these two star products both can be regarded as deformation quantization due to Weyl-Wigner correspondence \cite{Rieffel}.\\


\par More than constant commutation relation (\ref {2}), there may be defined other deformation structures on $\mathbb{R}^{2n}$ with linear and quadratic forms. The linear case leads to a Lie algebra with;
\begin{eqnarray}\label {3}
[x^\mu,x^\nu]=i\lambda_\sigma^{\mu\nu}~x^\sigma~,
\end{eqnarray}
\par \noindent $\lambda^{\mu\nu}_\sigma \in \mathbb{C}$. These structures are basically discussed in two different settings, fuzzy spaces \cite{Madore1, Madore2} and $\kappa$-deformation \cite{MR, DJMTWW}. The quadratic commutation relations are mostly given in terms of $R$-matrix or quasi-triangular structures of quantum groups \cite{RTF};
\begin{eqnarray} \label {444}
[x^\mu,x^\nu]=i(\frac{1}{q}R^{\mu\nu}_{\sigma\lambda}-\delta^\mu_\sigma~\delta^\nu_\lambda)~x^\sigma x^\lambda~,
\end{eqnarray}
\par \noindent $q\in \mathbb{C}$. $R$-matrices are the solutions of quantum Yang-Baxter equation in quantum inverse scattering theory \cite{Kassel}.\\
\par It is seen that the non-commutative structure (\ref {2}), despite of (\ref {3}) and (\ref {444}), is independent of coordinate functions $x^\mu$s in the chosen global coordinate chart $(\mathbb{R}^{2n},\{x^\mu\}_{\mu=1}^{2n})$. Such deformation structures are called translation-invariant products. More precisely a deformation quantization of $\mathbb{R}^m$ or a $\star$ product over $C^\infty(\mathbb{R}^m)$ is translation-invariant if and only if there exists a global coordinate chart on $\mathbb{R}^m$, say $(\mathbb{R}^{2n},\{x^\mu\}_{\mu=1}^{2n})$, such that (\ref {2}) holds for the coordinate functions. Translation-invariant products preserve the translation-invariance and consequently lead to the energy-momentum conservation law in substantially translation-invariant quantum field theories. More strictly, $\star$ is a translation-invariant star product on $C^\infty(\mathbb{R}^m)$, if and only if;
\begin{eqnarray} \label {5}
T_a (f)\star T_a (g)=T_a (f\star g)~,
\end{eqnarray}
\par \noindent for any vector $a\in \mathbb{R}^m$ and for any $f,g\in C^\infty (\mathbb{R}^m)$, where $T_a$ is the translating operator along a global coordinate system, say $\{x^\mu\}_{\mu=1}^m$; $T_a (f)(x)=f(x+a)$, $f\in C^\infty (\mathbb{R}^m)$. Replacing $a$ with $ta$, $t\in \mathbb{R}$, in (\ref {5}) and differentiating with respect to $t$ at $t=0$, one easily finds that $\partial_\mu (f\star g)=(\partial_\mu f)\star g+f\star (\partial_\mu g)$, $1\leq\mu\leq m$, which shows that any translation-invariant product is exact.\\
\par From now on for simplicity we suppose that there exists a fixed global coordinate chart over $\mathbb{R}^m$ with coordinate functions $\{x^\mu\}_{\mu=1}^m$. \\
\par An equivalent definition of translation-invariant products over Cartesian space $\mathbb{R}^m$, is given by \cite{GLV};
\begin{eqnarray} \label {6}
(f\star g)(x):=\int \frac{\emph{\emph{d}}^m p}{(2\pi)^m} \frac{\emph{\emph{d}}^m q}{(2\pi)^m} \tilde{f}(q)\tilde{g}(p)~e^{\alpha(p+q,q)} e^{i(p+q).x}~,
\end{eqnarray}
\par \noindent for $f,g\in C^\infty (\mathbb{R}^m)$, their Fourier transformations $\tilde{f},\tilde{g}$, and finally for generator $\alpha \in C^\infty (\mathbb{R}^m\times\mathbb{R}^m)$ which obeys the following cyclic property;
\begin{eqnarray} \label {7}
\alpha(p,r+s)+\alpha(r+s,r)=\alpha(p,r)+\alpha(p-r,s)~,
\end{eqnarray}
\par \noindent for any $p,r,s\in \mathbb{R}^m$. Actually (\ref {7}) is equivalent to associativity of generator $\star$, i.e. $(f\star g)\star h=f\star (g\star h)$ for $f,g,h\in C^\infty (\mathbb{R}^m)$. $\alpha$ is mostly referred to as the generator of $\star$. In fact, (\ref {6}) is the most general definition for translation-invariant deformation quantization of $\mathbb{R}^m$.\\
 \par To be precautious and to have well-defined products, from now on $C^\infty (\mathbb{R}^m)$ is replaced by $\mathcal{S}_c (\mathbb{R}^m)$, the Schwartz class functions with compactly supported Fourier transforms, for any translation-invariant product $\star$. It can be easily seen that $\mathcal{S}_c (\mathbb{R}^m)$ defines a well-defined closed algebra for any translation-invariant star product $\star$. The corresponding algebra is conventionally shown by $\mathcal{S}_c (\mathbb{R}^m)_\star$. Moreover, one can naturally extend the domain of the star products to $\mathcal{S}_{c,1} (\mathbb{R}^m ):=\mathcal{S}_c (\mathbb{R}^m)\bigoplus \mathbb{C}$ to have a unital algebra. The corresponding unital algebra with star product $\star$ is similarly shown by $\mathcal{S}_{c,1} (\mathbb{R}^m )_\star$. It is obvious that for any $f\in \mathcal{S}_{c,1} (\mathbb{R}^m)$, $1\star f=f\star 1=f$ if and only if;
 \begin{eqnarray} \label {8}
\alpha(p,p)=\alpha(p,0)=0~,
\end{eqnarray}
\par \noindent for any $p\in \mathbb{R}^m$. Combining (\ref {7}) and (\ref {8}) leads to;
\begin{eqnarray} \label {9}
\alpha(0,p)=\alpha(0,-p)~,
\end{eqnarray}
\par \noindent for any $p\in \mathbb{R}^m$. Using (\ref {9}) it can also be shown that any translation-invariant product admits the trace property;
\begin{eqnarray} \label {10}
\int_{\mathbb{R}^m} f_1\star \cdots \star f_{k-1}\star f_k =\int_{\mathbb{R}^m} f_k\star f_1\star \cdots \star f_{k-1}
\end{eqnarray}
for any $k\in \mathbb{N}$ and for any set of $f_1,\cdots,f_{k-1},f_k\in \mathcal{S}_c (\mathbb{R}^m)$. Also from (\ref {7})-(\ref {9}) it can be shown that;
\begin{eqnarray} \label {11}
	\alpha(p,q)=-\alpha(q,p)+\alpha(0,q-p)~,
\end{eqnarray}
\begin{eqnarray} \label {12}
\alpha(0,q)=\alpha(0,p)-\alpha(q,p)+\alpha(-p,q-p)~,
\end{eqnarray}
\par \noindent and finally
\begin{eqnarray} \label {13}
\alpha(p,q)=-\alpha(0,p)+\alpha(0,q)+\alpha(0,p-q)-\alpha(-p,q-p)~,
\end{eqnarray}
\par \noindent for any $p,q\in \mathbb{R}^m$. On the other hand, it is easily seen that the commutivity of $\star$ is equivalent to;
\begin{eqnarray} \label {14}
\alpha(p,q)=\alpha(p,p-q)~,
\end{eqnarray}
\par \noindent for any $p,q\in \mathbb{R}^m$. Thus $\alpha$ is called commutative if it satisfies (\ref {14}).\\


\par
\section{$\alpha$-Cohomology and Classification of Translation-Invariant Star Products}
\setcounter{equation}{0}
\par  Let $\emph{\emph{C}}^n (\mathbb{R}^m )\subseteq C^\infty (\underbrace{\mathbb{R}^m\times\cdots \times\mathbb{R}^m}_{n-\emph{\emph{fold}}})$, $n\geq0$, be the complex vector spaces generated by smooth functions $f$ with properties of:
\begin{itemize}
  \item $\emph{\emph{C}}^0(\mathbb{R}^m):=\{0\}$,
  \item For $n=1$; $\emph{\emph{C}}^1(\mathbb{R}^m):=\{f\in C^\infty(\mathbb{R}^m)|f(0)=0\}$,
  \item For $n=2$; $\emph{\emph{C}}^2(\mathbb{R}^m):=\{f\in C^\infty(\mathbb{R}^m\times \mathbb{R}^m)|f(p,0)=f(p,p)=0; p\in \mathbb{R}^m\}$,
  \item For $n\geq3$; $\emph{\emph{C}}^n(\mathbb{R}^m)\subseteq C^\infty(\underbrace{\mathbb{R}^m\times ...\times\mathbb{R}^m}_{n- \emph{\emph{fold}}})$ consists of smooth functions $f$ with properties of $f(p_1,...,p_{n-1},0)=f(p_1,...,p_k,p,p,p_{k+1},...,p_{n-2})=0$, $k\leq {n-2}$, for any $p,p_1,...,p_{n-1}\in \mathbb{R}^m$,\\
\end{itemize}
\par \noindent then consider the linear maps
\begin{eqnarray} \label {15}
\partial_n:\emph{\emph{C}}^n(\mathbb{R}^m)\longrightarrow \emph{\emph{C}}^{n+1}(\mathbb{R}^m)~,
\end{eqnarray}
\par \noindent commonly denoted by $\partial$, defined by;
\begin{eqnarray} \label {16}
\partial_n f(p_0,...,p_n):=\varepsilon_n \sum _{i=0}^n f(p_0,...,p_{i-1},\hat{p _i},p_{i+1},...,p_n)+\varepsilon_n (-)^{n+1} f(p_0-p_n,\cdots,p_{n-1}-p_n)~,
\end{eqnarray}
\par \noindent $f\in\emph{\emph{C}}^n(\mathbb{R}^m)$, with $\varepsilon_n=1$ for odd $n$ and $\varepsilon_n=i$ for $n$ even. It can be seen that ; $\partial^2=\partial_n\circ\partial_{n-1}=0$ for any $n\in \mathbb{N}$. Therefore, $\emph{\emph{C}}^n (\mathbb{R}^m )$s as cochains and $\partial_n$s as coboundary maps define a complex with;
\begin{eqnarray} \label {17}
\emph{\emph{C}}^0(\mathbb{R}^m)
\begin{array}{c}
  {\partial_0} \\
  {\longrightarrow} \\
  { }
\end{array}
\emph{\emph{C}}^1(\mathbb{R}^m)
\begin{array}{c}
  {\partial_1} \\
  {\longrightarrow} \\
  { }
\end{array}
...
\begin{array}{c}
  {\partial_{n-1}} \\
  {\longrightarrow} \\
  { }
\end{array}
\emph{\emph{C}}^n(\mathbb{R}^m)
\begin{array}{c}
  {\partial_{n}} \\
  {\longrightarrow} \\
  { }
\end{array}
... ~.
\end{eqnarray}

\par The complex (\ref {17}) defines a cohomology theory so called $\alpha$-cohomology \cite{GLV, Galluccio, Varshovi} which is deduced from Hochschild cohomology. As a generic convention the notation of $\alpha_1\sim\alpha_2$ is used for two $\alpha$-cohomologous $n$-cocycles $\alpha_1$ and $\alpha_2$. Also the cohomology class of $\alpha\in Ker\partial_n$ is shown by $[\alpha]$. Therefore, the $\alpha$-cohomolgy group, $H_\alpha^n(\mathbb{R}^m):=Ker\partial_n/Im\partial_{n-1}$, classifies $n$-cocycles differing in coboundary terms into the same equivalence classes. Now consider the translation-invariant products given by $\alpha\in \emph{\emph{C}}^2(\mathbb{R}^m)$ due to definition (\ref {6}). According to (\ref {7}), associativity of $\star$ is equivalent to $\partial \alpha=0$. More precisely, $\alpha$ is a generator if and only if it is a 2-cocycle. Indeed, $H_\alpha^2 (R^m)$ classifies all the translation-invariant quantization structures of $\mathcal{S}_{c,1}(\mathbb{R}^m)$ modulo the coboundary terms. It can be easily seen from (\ref {14}) that if $[\alpha]=0$ then $\alpha$ is commutative. In the following it is shown that the inverse is also true and thus $H_\alpha^2 (\mathbb{R}^m)$ classifies the translation-invariant star products on $\mathcal{S}_{c,1}(\mathbb{R}^m)$ modulo the commutative ones. To see this fact set;
\begin{eqnarray} \label {18}
\alpha^{'}(p,q):=\frac{1}{2} (\alpha(p,q)+\alpha(-p,-q))~,
\end{eqnarray}
\par \noindent for $p,q\in \mathbb{R}^m$. It can be checked that $\partial\alpha^{'}=0$ and thus it defines a translation-invariant structure on $\mathcal{S}_{c,1} (\mathbb{R}^m)$. Next define $\star^{'}$ with $\alpha^{'}$ accordingly;
\begin{eqnarray} \label {19}
(f\star^{'} g)(x):=\int \frac{\emph{\emph{d}}^m p}{(2\pi)^m}\frac{\emph{\emph{d}}^m q}{(2\pi)^m} \tilde{f}(q) \tilde{g}(p-q)~e^{\alpha^{'}(p,q)}~ e^{ip.x}~.
\end{eqnarray}
\par \noindent for $f,g\in \mathcal{S}_{c,1}(\mathbb{R}^m)$.\\
\par Now let $\alpha^{''}:=\alpha-\alpha^{'}$. More precisely;
\begin{eqnarray} \label {20}
\alpha^{''} (p,q)=\frac{1}{2} (\alpha(p,q)-\alpha(-p,-q))~.
\end{eqnarray}
\par \noindent for any $p,q\in \mathbb{R}^m$. Therefore, $\partial\alpha^{''}=0$. Using (\ref {14}) it can be shown that $\alpha^{''}$ is commutative. Thus, $\star$ and $\star^{'}$ differ in an associative commutative product. On the other hand; $\alpha^{''}(p,q)=-\alpha^{''}(-p,-q)$ for any $p,q\in \mathbb{R}^m$. Then, (\ref {9}) and (\ref {11}) lead to;
\begin{eqnarray} \label {21}
\alpha^{''}(p,q)=-\alpha^{''}(q,p)
\end{eqnarray}
\par \noindent for $p,q\in \mathbb{R}^m$. Indeed $\alpha^{''}$ obeys the following properties;
\begin{eqnarray} \label {22}
\left\{
  \begin{array}{ll}
     \alpha^{''}(p,q)=\alpha^{''}(p,p-q)~\\
     \alpha^{''}(p,q)=-\alpha^{''}(-p,-q)~,\\
     \alpha^{''}(p,q)=-\alpha^{''}(q,p)~
  \end{array}
\right.
\end{eqnarray}\\
\noindent for any $p,q\in \mathbb{R}^m$. In appendix A it is shown that (\ref {22}) and $\partial\alpha^{''}=0$ give hand an element of $\emph{\emph{C}}^1 (\mathbb{R}^m)$, say $\beta$, which $\alpha^{''}=\partial\beta$. In fact, $\alpha^{''}$ is a coboundary and thus; $\alpha \sim \alpha^{'}$.\\
\par Now we are ready to show that $\alpha$ is commutative if and only if $\alpha$ is a coboundary. As mentioned above the only if part is obvious, so it is sufficient to prove the if term. According to appendix A, to prove this fact, it is enough to show that $\alpha^{'}$ is also a coboundary provided $\alpha$ is commutative. Using (\ref {13}) for commutative $\alpha$, (\ref {18}) leads to;
\begin{eqnarray} \label {23}
\alpha^{'}(p,q)=\frac{1}{2} (\alpha(0,q)-\alpha(0,p)+\alpha(0,p-q))=\frac{1}{2} \partial\alpha_0(p,q)~,
\end{eqnarray}
\par \noindent for $\alpha_0 (p)=\alpha(0,p)\in \emph{\emph{C}}^1 (\mathbb{\mathbb{R}}^m)$ and for any $p,q\in \mathbb{R}^m$. This shows that $\alpha_1 \sim \alpha_2$ if and only if $\alpha_1-\alpha_2$ is commutative. Consequently:\\

 \par \textbf{Theorem 1;} \emph{ $H_\alpha^2(\mathbb{R}^m)$ classifies the translation-invariant star products on $\mathcal{S}_{c,1}(\mathbb{R}^m)$ modulo the commutative ones.}\\

 \par As an example the Groenewold-Moyal product, $\star_{G-M}$, and the Wick-Voros production, $\star_{W-V}$, are respectively generated by 2-cocycles $\alpha_{G-M}(p,q)=iq^\mu \theta_{A~\mu\nu}p^\nu$ and $\alpha_{W-V}(p,q)=\alpha_{G-M}(p,q)+q^\mu \theta_{S~\mu\nu}(p-q)^\nu$, with $p,q\in \mathbb{R}^m$, for $\theta_S$ ($\theta_A$) a fixed (anti-) symmetric real matrix. It is clear that $\alpha_{G-M}$ and $\alpha_{W-V}$ differ in a commutative generator. Therefore, $\alpha_{G-M}$ and $\alpha_{W-V}$ are $\alpha$-cohomologous and belong to the same class of $H_\alpha^2 (\mathbb{R}^m)$, commonly denoted by $[\alpha_{G-M}]$.\\


\par
\section{The Hodge Theorem in $\alpha$-Cohomology}
\setcounter{equation}{0}

\par In (\ref {18}) one corresponds to each 2-cocycle $\alpha$ an $\alpha$-cohomologous element $\alpha^{'}$ with property
\begin{eqnarray} \label {24}
\alpha^{'} (p,q)=\alpha^{'} (-p,-q)~,
\end{eqnarray}
\par \noindent for any $p,q\in \mathbb{R}^m$. It can also be seen that
\begin{eqnarray} \label {25}
\alpha(p,q)=\alpha_-(p,q)+\alpha_+(p,q)
\end{eqnarray}
with
\begin{eqnarray} \label {26}
\begin{array}{c}
  \alpha_- (p,q):=\frac{1}{2} ( \alpha(p,q)-\alpha(-p,q-p)  )~,\\ \\
  \alpha_+ (p,q):=\frac{1}{2} ( \alpha(p,q)+\alpha(-p,q-p)  )~,
\end{array}
\end{eqnarray}
\par \noindent defines another such correspondence for 2-cocycle $\alpha$. By (\ref {13}) it can be seen that $\alpha_+=\frac{1}{2} \partial\alpha_0$, for $\alpha_0$ the 1-cochain defined in (\ref {23}), and thus $\alpha$ and $\alpha_-$ define two $\alpha$-cohomologous 2-cocycles. In fact, $\alpha_-$ satisfies the condition
\begin{eqnarray} \label {27}
\alpha_- (p,q)=-\alpha_- (-p,q-p)~,
\end{eqnarray}
\par \noindent for any $p,q\in \mathbb{R}^m$. Therefore, according to (\ref {18}) and (\ref {25}), one can correspond to each 2-cocycle $\alpha$ an $\alpha$-cohomologous element $\alpha_-^{'}:=(\alpha^{'})_-=(\alpha_-)^{'}$ with
\begin{eqnarray} \label {28}
\alpha_-^{'} (p,q)=\alpha_-^{'} (-p,-q)=-\alpha_-^{'} (-p,q-p)~.
\end{eqnarray}
\par It is claimed that such correspondence is unique. That is for any cohomology class $[\alpha]\in H_\alpha^2 (\mathbb{R}^m)$ there is a unique element, $\alpha_-^{'}\in[\alpha]$, satisfying (\ref {28}). This can be considered as the Hodge theorem \cite{Warner} for $\alpha$-cohomology classes of $H_\alpha^2 (\mathbb{R}^m)$, since then any given 2-cocycle $\alpha$ can be uniquely decomposed to: $\alpha=\alpha_-^{'}+\partial \beta$, for unique $\alpha_-^{'}$ satisfying (\ref {28}) and unique coboundary term $\partial \beta$. Indeed $\alpha_-^{'}$ is the unique solution of
\begin{eqnarray} \label {29}
\Delta\alpha(p,q):=\alpha(0,q)-\alpha(0,p)+\alpha(0,p-q)+\alpha(p,q)+\alpha(p,p-q)=0~,
\end{eqnarray}
\par \noindent $p,q\in \mathbb{R}^m$, in $[\alpha]\in H_\alpha^2 (\mathbb{R}^m)$. Here $\Delta$ can be considered as a Laplace-Beltrami operator on the cochain $\emph{\emph{C}}^2 (\mathbb{R}^m)$ of complex (\ref {17}). Therefore, $\alpha_-^{'}$ is called the harmonic element or the harmonic form of $[\alpha]$. Similarly a harmonic translation-invariant product is a translation-invariant product generated by a harmonic form. A collection of manipulations shows that
\begin{eqnarray} \label {30}
\alpha_-^{'} (p,q)=\frac{1}{2}(\alpha(p+q,q)-\alpha(p+q,p))~,
\end{eqnarray}
\par \noindent which can be checked directly by (\ref {28}) with considering the uniqueness of such elements in each $\alpha$-cohomology class. More precisely, for any given 2-cocycle $\alpha$ its $\alpha$-cohomologous harmonic form can be directly calculated from (\ref {30}). Moreover it is not difficult to see that
\begin{eqnarray} \label {31}
\alpha_-^{'} (p+nq,q)=\alpha_-^{'} (p,q)~,
\end{eqnarray}
\par \noindent for any $p,q\in \mathbb{R}^m$ and for any $n\in \mathbb{Z}$. Equations (\ref {30}) and (\ref {31}) will be proven directly in section $\emph{\emph{VIII}}$. To prove the Hodge theorem for $\alpha$-cohomology classes consider two $\alpha$-cohomologous harmonic 2-cocycles $\alpha_1$ and $\alpha_2$. It is enough to show that; $\alpha_1=\alpha_2$. To see this let $\partial\beta=\alpha_1-\alpha_2$ for some $\beta\in \emph{\emph{C}}^1(\mathbb{R}^m)$. Thus for any $p,q\in \mathbb{R}^m$, (\ref {28}) leads to;
\begin{eqnarray} \label {32}
\beta(-q)-\beta(-p)+\beta(q-p)=-(\beta(q-p)-\beta(-p)+\beta(-q))~,
\end{eqnarray}
\par \noindent and hence; $\partial\beta=0$. This proves the claim. According to (\ref {28}), (\ref {11}) and (\ref {13}), any harmonic form $\alpha$ of an $\alpha$-cohomology class, obeys the properties of;
\begin{eqnarray} \label {33}
\left\{
  \begin{array}{ll}
    \alpha(p,q)=-\alpha(p,p-q) \\
    \alpha(p,q)=\alpha(-p,-q)~~, \\
    \alpha(p,q)=-\alpha(q,p)
  \end{array}
\right.
\end{eqnarray}
\noindent for any $p,q\in \mathbb{R}^m$. Therefore, we have just proven the following theorem;\\

\par \textbf{Theorem 2;} \emph{Any given 2-cocycle $\alpha$ can be uniquely decomposed to $\alpha=\alpha_H+\partial\beta$, where $\alpha_H$, the harmonic form, is the unique element of $\alpha$-cohomology class $[\alpha]$ which satisfies the conditions (\ref {33}).}\\

 \par In the following this theorem will be referred to as \emph{the Hodge theorem} for $\alpha$-cohomology. The most important property of harmonic forms is cleared in integration of star product of functions. In fact, it can be shown that for $\star$, a harmonic translation-invariant product, one finds that;
\begin{eqnarray} \label {34}
\int_{\mathbb{R}^m} f\star g=\int_{\mathbb{R}^m} fg
\end{eqnarray}\\
\noindent for any $f,g\in \mathcal{S}_c (\mathbb{R}^m)$. To see this note that;
\begin{eqnarray} \label {35}
\int_{\mathbb{R}^m} f\star g=\int \frac{\emph{\emph{d}}^m p}{(2\pi)^m}  \emph{\emph{d}}^m x  \tilde{f}(p)\tilde{g}(-p) e^{\alpha(0,p)}=\int_{\mathbb{R}^m}fg
\end{eqnarray}\\
\noindent for $f,g\in \mathcal{S}_c (\mathbb{R}^m)$, since $\alpha(0,p)=0$, $p\in\mathbb{R}^m$, for any harmonic form $\alpha$ due to (\ref {8}) and (\ref {33}). For instance by considering (\ref {33}) it easy to see that $\alpha_{G-M}$ is the harmonic form of $\alpha$-cohomology class $[\alpha_{G-M}]$ and thus $\star_{G-M}$ satisfies the condition (\ref {35}).\\


\par
\section{$\alpha^*$-Cohomology for Complex Translation-Invariant Products}
\setcounter{equation}{0}

\par The other prominent property of harmonic forms is cleared in the setting of $\alpha^*$-cohomology theory. $\alpha^*$-cohomology is defined by complex;


\begin{eqnarray} \label {36}
0=\emph{\emph{C}}_*^0(\mathbb{R}^m)
\begin{array}{c}
  {\partial_0} \\
  {\longrightarrow} \\
  { }
\end{array}
\emph{\emph{C}}_*^1(\mathbb{R}^m)
\begin{array}{c}
  {\partial_1} \\
  {\longrightarrow} \\
  { }
\end{array}
...
\begin{array}{c}
  {\partial_{n-1}} \\
  {\longrightarrow} \\
  { }
\end{array}
\emph{\emph{C}}_*^n(\mathbb{R}^m)
\begin{array}{c}
  {\partial_{n}} \\
  {\longrightarrow} \\
  { }
\end{array}
...
\end{eqnarray}


\par \noindent where $\emph{\emph{C}}_*^n (\mathbb{R}^m )$ is the cochain of elements $f\in \emph{\emph{C}}^n (\mathbb{R}^m)$, with property
\begin{eqnarray} \label {37}
f^* (p_1,...,p_n )=f(-p_1,p_n-p_1,p_{n-1}-p_1,...,p_2-p_1)~,
\end{eqnarray}
\par \noindent for $f^*$ the complex conjugate of $f$ and for any collection of $p_1,...,p_n\in \mathbb{R}^m$. The elements of $Ker\partial_n$ and $Im\partial_{n-1}$ in complex (\ref {36}) are respectively called complex $n$-cocycles and complex $n$-coboundaries. Conventionally $H_{\alpha^*}^n(\mathbb{R}^m)$ is used for the $n$th cohomology group of complex (\ref {36}), or more precisely for the $n$th $\alpha^*$-cohomology group. Moreover, it can be easily shown that the complex inclusion map

\begin{eqnarray} \label {38}
\begin{array}{ccccccccc}
  { } & {\partial_0} &   & {\partial_1} & {~}  & {\partial_{n-1}} & { } & {\partial_n} & { } \\
  {0=\emph{\emph{C}}_*^0(\mathbb{R}^m)} & {\longrightarrow} & {\emph{\emph{C}}_*^1(\mathbb{R}^m)} & {\longrightarrow} & {...} & {\longrightarrow} & {0=\emph{\emph{C}}_*^n(\mathbb{R}^m)} & {\longrightarrow} & {...} \\
  {i_0\downarrow} & {\partial_0} & {i_1\downarrow} & {\partial_1} & {...} & {\partial_{n-1}} & {i_n\downarrow} & {\partial_n} & {...} \\
  {0=\emph{\emph{C}}^0(\mathbb{R}^m)} & {\longrightarrow} & {\emph{\emph{C}}^1(\mathbb{R}^m)} & {\longrightarrow} & {...} & {\longrightarrow} & {\emph{\emph{C}}^n(\mathbb{R}^m)} & {\longrightarrow} & {...}
\end{array}
\end{eqnarray}\\


\noindent for inclusions $i_n:\emph{\emph{C}}_*^n (\mathbb{R}^m )\rightarrow \emph{\emph{C}}^n (\mathbb{R}^m)$, leads to a family of inclusions of cohomology groups;
\begin{eqnarray} \label {39}
{i_n}_*:H_{\alpha^*}^n (\mathbb{R}^m)\hookrightarrow H_\alpha^n (\mathbb{R}^m)~.
\end{eqnarray}


\par Consequently, the $n$th group of $\alpha^*$-cohomology, $H_{\alpha^*}^n(\mathbb{R}^m)$, is a subgroup of the $n$th group of $\alpha$-cohomology, $H_{\alpha}^n(\mathbb{R}^m)$. Therefore, it seems that by studying the $\alpha^*$-cohomology we focus our attention to a special kind of star products. In fact, by this restriction we are looking for those star products which obey the following property;
\begin{eqnarray} \label {40}
(f\star g)^*=g^*\star f^*~,
\end{eqnarray}
\par \noindent for any $f,g\in \mathcal{S}_{c,1} (\mathbb{R}^m)$. Such multiplications are mostly referred to as complex products or involution structures. Since from the view points of Lagrangian formalism the action must be a real number, complex products are the only suitable multiplications in order to be used in quantum physics, where the fields can be imaginary in general. Therefore, as it will be cleared in the following, to illuminate the quantum features of translation-invariant star products, studying the theory of $\alpha^*$-cohomology comes more effective.\\


 \par To see this fact recall that according to (\ref {37}) any complex 2-cocycle $\alpha$ satisfies the condition
\begin{eqnarray} \label {41}
\alpha^* (p,q)=\alpha(-p,q-p)~,
\end{eqnarray}
\par \noindent for any $p,q\in \mathbb{R}^m$. On the other hand, it can also be seen that the generator of any complex translation-invariant star product satisfies the condition (\ref {41}). Thus, the group $H_{\alpha^*}^2 (\mathbb{R}^m)$ classifies complex translation-invariant deformations of $\mathcal{S}_{c,1} (\mathbb{R}^m)$ up to commutative complex products. Moreover, since; $\partial_n f^*=\pm(\partial_n f)^*$, and $f^*\in \emph{\emph{C}}_*^n (\mathbb{R}^m)$ for any $f\in \emph{\emph{C}}_*^n (\mathbb{R}^m)$, then $\alpha^*\in \emph{\emph{C}}_*^2 (\mathbb{R}^m)$ is a complex 2-cocycle when $\alpha$ is a complex 2-cocycle. Therefore, to any $\alpha^*$-cohomology class, say $[\alpha]\in H_{\alpha^*}^2 (\mathbb{R}^m)$, one can correspond a conjugate $\alpha^*$-cohomology class by $[\alpha^*]$. The condition (\ref {41}) together with (\ref {13}) asserts that if $[\alpha]$ belongs to $H_{\alpha^*}^2 (\mathbb{R}^m)\subseteq H_\alpha^2 (\mathbb{R}^m)$, then $[\alpha^*]$ is the dual of $[\alpha]$ in the sense of;
\begin{eqnarray} \label {42}
[\alpha]+[\alpha^* ]=0~.
\end{eqnarray}


 \par This is the pure imaginary condition for $\alpha$-cohomology classes. On the other hand, since we have ${\alpha_-^{'}}^*(p,q)=\alpha_-^{'}(-p,q-p)$ for any $\alpha\in \emph{\emph{C}}_*^2 (\mathbb{R}^m )$ and for any $p,q\in \mathbb{R}^m$, it is seen that \emph{the Hodge theorem} is also true for $H_{\alpha^*}^2 (\mathbb{R}^m)$, and therefore any harmonic form $\alpha$ of an $\alpha^*$-cohomology class of $H_{\alpha^*}^2 (\mathbb{R}^m)$, is pure imaginary;
\begin{eqnarray} \label {43}
\alpha^*=-\alpha~.
\end{eqnarray}
\par Indeed, since we have ${\alpha_-^{'}}^*=-\alpha_-^{'}$ for $\alpha$ a complex 2-cocycle of $\emph{\emph{C}}_*^2 (\mathbb{R}^m )$, then (\ref {43}) follows directly. Conversely, by using the complex conjugate of (\ref {33}), it can be seen that any harmonic form $\alpha$ defines an $\alpha^*$-cohomology class in $H_{\alpha^*}^2 (\mathbb{R}^m)$, if it is pure imaginary. Consequently, this asserts that $H_{\alpha^*}^2 (\mathbb{R}^m)$ is the collection of all pure imaginary classes of $H_\alpha^2 (\mathbb{R}^m)$. Moreover, (\ref {33}) says that $\alpha$ is a harmonic form if and only if $\alpha^*$ is harmonic. Therefore, by replacing $\alpha$ with $1/2 (\alpha+\alpha^*)+1/2 (\alpha-\alpha^*)$, one finds; $\emph{\emph{dim}}~H_\alpha^2 (\mathbb{R}^m)=2 ~\emph{\emph{dim}}~H_{\alpha^*}^2 (\mathbb{R}^m)$. Actually, the decomposition of $\alpha=Re(\alpha)+iIm(\alpha)$ for any harmonic 2-cocycle $\alpha$ of $\emph{\emph{C}}^2 (\mathbb{R}^m)$ shows that;\\

\par \textbf{Theorem 3;} \emph{$H_\alpha^2 (\mathbb{R}^m )=H_{\alpha^*}^2 (\mathbb{R}^m)\bigotimes_\mathbb{R} \mathbb{C}$. More precisely, $H_{\alpha^*}^2 (\mathbb{R}^m)$ consists of all pure imaginary elements of $H_\alpha^2 (\mathbb{R}^m )$.} \\


 \par For instance, 2-cocycles $\alpha_{G-M}$ and $\alpha_{W-V}$ both satisfy the condition (\ref {41}) and consequently define two complex translation-invariant products on $\mathcal{S}_{c,1} (\mathbb{R}^m)$. Moreover, since $\alpha_{G-M}$ is a harmonic complex 2-cocycle, then it must be pure imaginary due to (\ref {43}).\\
 \par It would be interesting to study the role of the second group of $\alpha^*$-cohomology in classifying the algebraic complex structures on $\mathcal{S}_{c,1} (\mathbb{R}^m)$ due to translation-invariant products. By definition two complex translation-invariant products $\star_1$ and $\star_2$ are equivalent, and then we write $\star_1\sim\star_2$, if there exists an invertible complex translation-invariant linear map $T:\mathcal{S}_{c,1} (\mathbb{R}^m)\rightarrow \mathcal{S}_{c,1} (\mathbb{R}^m)$ such that for any two elements $f,g\in \mathcal{S}_{c,1} (\mathbb{R}^m)$;
\begin{eqnarray} \label {44}
T(f\star_1 g)=T(f) \star_2 T(g)~.
\end{eqnarray}

\par Strictly speaking $T$ is an algebra isomorphism from $\mathcal{S}_{c,1} (\mathbb{R}^m)_{\star_1}$ to $\mathcal{S}_{c,1} (\mathbb{R}^m)_{\star_2}$ with $T(f^* )=T(f)^*$ and $T(\partial_\mu f)=\partial_\mu T(f)$ for $f\in \mathcal{S}_{c,1} (\mathbb{R}^m)$ and $1\leq\mu\leq m$. It can be shown that $\star_1\sim\star_2$ if and only if $\alpha_1\sim \alpha_2$. To see this, according to (\ref {44}) assume that $\star_1\sim\star_2$ with $T=e^\triangle$ for $\triangle$ a translation-invariant differential operator with respect to coordinate functions $x^\mu$, $1\leq\mu\leq m$. Thus, using integration by part one can show that;
\begin{eqnarray} \label {45}
\widetilde{T(f)}(p)=\int \emph{\emph{d}}^m x e^{-ip.x} e^\triangle f =e^{\tilde{\triangle}(p)} \tilde{f}(p)~,
\end{eqnarray}
\par \noindent for any $p,q\in \mathbb{R}^m$ and for any $f\in S_{c,1} (\mathbb{R}^m)$.\\


 \par Since $T$ is an algebra isomorphism then; $T(1)=1$, and consequently it can be seen that; $\tilde \triangle(0)=0$. On the other hand, $T$ is complex, i.e.; $T(f^* )={T(f)}^*$ for any $f\in \mathcal{S}_{c,1} (\mathbb{R}^m)$, therefore one finds; ${\tilde{\triangle}}^*(p)=\tilde{ \triangle}(-p)$. Hence; $\tilde{\triangle}\in \emph{\emph{C}}_*^1 (\mathbb{R}^m)$. Now one can compute;
\begin{eqnarray} \label {46}
\begin{array}{c}
  {(T(f)\star_2 T(g))(x)} \\\\
  {=\int \frac{\emph{\emph{d}}^m p}{(2\pi)^m} \frac{\emph{\emph{d}}^m q}{(2\pi)^m}~\widetilde{T(f)(q)}~\widetilde{T(g)}(p-q)~e^{\alpha_2(p,q)}~e^{ip.x}} ~~~~~~~\\\\
  {=\int \frac{\emph{\emph{d}}^m p}{(2\pi)^m} \frac{\emph{\emph{d}}^m q}{(2\pi)^m}~\tilde{f}(q)~\tilde{g}(p-q)~} e^{\alpha_2(p,q)+\tilde{\triangle}(q)+\tilde{\triangle}(p-q)}~e^{ip.x} \\\\
  {=\int \frac{\emph{\emph{d}}^m p}{(2\pi)^m} \frac{\emph{\emph{d}}^m  q}{(2\pi)^m}~\tilde{f}(q)~\tilde{g}(p-q)~e^{\alpha_2(p,q)+\partial\tilde{\triangle}(p,q)+\tilde{\triangle}(p)}~e^{ip.x}} \\\\
  {=\int \frac{\emph{\emph{d}}^m p}{(2\pi)^m}~\widetilde{T(f\star g)}(p)~e^{ip.x}}~~~~~~~~~~~~~~~~~~~~~~~~~~~~~~~~~~~~ \\\\
  =T(f\star g)(x)
\end{array}
\end{eqnarray}


\par \noindent for any $f,g\in \mathcal{S}_{c,1} (\mathbb{R}^m)$, and for $\star$ the translation-invariant product generated by $\alpha_2+\partial\tilde{\triangle}$. Thus (\ref {44}) asserts that; $\alpha_1=\alpha_2+\partial\tilde{\triangle}$, and consequently; $\alpha_1\sim\alpha_2$. Now, for the converse assume that $\alpha_1=\alpha_2+\partial\tilde{\triangle}$ for $\partial\tilde{\triangle}\in \emph{\emph{C}}_*^1(\mathbb{R}^m)$. By (\ref {44}) and (\ref {46}) it is enough to define; $T:=e^\triangle$. This shows that $H_{\alpha^*}^2 (\mathbb{R}^m)$ classifies all *-algebraic (complex) translation-invariant structures of $\mathcal{S}_{c,1} (\mathbb{R}^m)$ up to isomorphism. Generally by forgetting the complex structures it can be similarly shown that $H_\alpha^2 (\mathbb{R}^m)$ also classifies all the algebraic structures of $\mathcal{S}_{c,1} (\mathbb{R}^m)$ due to translation-invariant products up to isomorphism. Therefore, we have just shown that;\\

\par \textbf{Theorem 4;} \emph{Two (complex) translation-invariant star products are equivalent if and only if their generators are ($\alpha^*$-) $\alpha$-cohomologous. More precisely, ($H_{\alpha^*}^2(\mathbb{R}^m)$) $H_{\alpha}^2(\mathbb{R}^m)$ classifies all (*-) algebraic translation-invariant structures of $\mathcal{S}_{c,1}(\mathbb{R}^m)$ up to isomorphism.}


\par
\section{Classification of Translation-Invariant Quantum Field Theories}
\setcounter{equation}{0}

\par To study the complex translation-invariant products more exhaustively, their role should be discussed during the standard loop calculations for general translation-invariant non-commutative quantum field theories. It can be seen that the $\alpha^*$-cohomology class of a complex translation-invariant product $\alpha$, solely describes the UV/IR mixing behavior of translation-invariant $\phi_{\star}^4$-theory \cite{GLV} up to one loop corrections. Actually, the non-planar corrections to 2-point function in $\phi_{\star}^4$-theory on $\mathbb{R}^4$ at one-loop level are given by;
\begin{eqnarray} \label {47}
\int \frac{\emph{\emph{d}}^4 q}{(2\pi)^4}~\frac{e^{-\alpha(0,p)+\omega_\alpha(p,q)}}{(p^2-m^2)^2~(q^2-m^2)}~,
\end{eqnarray}
\par \noindent $p\in \mathbb{R}^4$, with;
\begin{eqnarray} \label {48}
\omega_\alpha(p,q):=\alpha(p+q,p)-\alpha(p+q,q)~
\end{eqnarray}
\par \noindent for any $p,q\in \mathbb{R}^4$. It is easily seen by (\ref {14}) that $\omega_\alpha=0$ if and only if $\alpha$ is commutative. This lets one to define $\omega$ on classes of $H_{\alpha^*}^2 (\mathbb{R}^m)$. More strictly, according to (\ref {30}) $\omega_\alpha$ is $-2$ times of $\alpha_-^{'}$. In fact, defining $\omega$ on $H_{\alpha^*}^2 (\mathbb{R}^m)$ classifies the UV/IR mixing of translation-invariant $\phi_\star^4$-theory at one loop corrections. It is also known \cite{TV} that such classification comes true in all finite orders of perturbation for 2- and 4-point functions.\\


\par In the following, using the concept of harmonic forms, a simple non-perturbative proof is given for the possibility of classifying the quantum behaviors of translation-invariant non-commutative quantum field theories by means of $\alpha^*$-cohomology. It is seen that our proof is exact; i.e. it includes all loop corrections and all $n$-point functions. To see this, consider two different versions of an arbitrary quantum field theory due to two equivalent translation-invariant complex products $\star_1$ and $\star_2$. Assume that the theory contains $n$ fields $\{\phi_i\}_{i=1}^n$. Consider the most general interacting action;


\begin{eqnarray} \label {49}
S_\star^{int}=\int_{\mathbb{R}^m}~\phi_{i_1,(\mu_{i_1})}...\phi_{i_k,(\mu_{i_k})}~,
\end{eqnarray}


\par \noindent where $\phi_{,(\mu)}=\partial_{(\mu)} \phi$ for multi-index $(\mu)$. The star product $\star$ in (\ref {49}) stands for $\star_1$ and $\star_2$, accordingly. For the case of $\star=\star_1$, the interaction term (\ref {49}) can be rewritten in the phase space with;
\begin{eqnarray} \label {50}
 S_{\star_1}^{int}=\int~\prod_{j=1}^k~\frac{\emph{\emph{d}}^m p^j}{(2\pi)^m}~p_{(\mu_{i_1})}^1~\tilde{\phi_{i_1}}(p^1)~~...~~p_{(\mu_{i_k})}^k~\tilde{\phi_{i_k}}(p^k)~V_{\star_1}^k (p^1,...,p^k )~\delta^{(m)}~ (\sum_{j=1}^k p^j)
\end{eqnarray}


\par \noindent for

\begin{eqnarray} \label {51}
V_{\star}^k (p^1,...,p^k )=e^{\sum_{i=2}^k\alpha (\sum_{j=1}^ip^j ,\sum_{j=1}^{i-1}p^j)}
\end{eqnarray}


\par \noindent the non-commutative vertex for translation-invariant star product $\star$ generated by $\alpha$. According to (\ref {45}), the redefinition of $\phi_i^{'}=e^\delta~\phi_i$, $i=1,...,n$, gives $S_{\star_1}^{int}$ in terms of $V_{\star_2}^k$ with;
\begin{eqnarray} \label {52}
 S_{\star_1}^{int}=\int~\prod_{j=1}^k~\frac{\emph{\emph{d}}^m p^j}{(2\pi)^m}~p_{(\mu_{i_1})}^1~\tilde{\phi_{i_1}^{'}}(p^1)~~...~~p_{(\mu_{i_k})}^k~\tilde{\phi_{i_k}^{'}}(p^k)~V_{\star_2}^k (p^1,...,p^k )~\delta^{(m)}~ (\sum_{j=1}^k p^j)~.
\end{eqnarray}\\
 \par More precisely;
\begin{eqnarray} \label {53}
S_{\star_1}^{int} (\phi_1,...,\phi_n )=S_{\phi_2}^{int} (\phi_1^{'},...,\phi_n^{'})~.
\end{eqnarray}


  \par Thus, it is enough to show that the propagators of $\phi_i^{'}$s, $i=1,...,n$, are also given in terms of $\alpha_2$, the generator of $\star_2$. It is not hard to see that without loss of generality one can restrict the issue to scalar or Klein-Gordon like fields. For vector and spinor fields the same result can be deduced similarly. Here, for a while we restrict ourselves to scalar Klein-Gordon like fields, and then we will back soon to the general case. For Klein-Gordon like fields the propagator $\widetilde{G_{\star_1}^\phi}$ for the quantum field $\phi$ is given by
\begin{eqnarray} \label {54}
\widetilde{\Xi_{\star_1}^\phi(p)}~\widetilde{G_{\star_1}^\phi}(p)=1
\end{eqnarray}
\par \noindent for
\begin{eqnarray} \label {55}
 S_{\star_1}^0=\sum_{i=1}^n\int \frac{\emph{\emph{d}}^m p}{(2\pi)^m}~\frac{1}{2} \tilde{\phi_i}(p)~\tilde{\phi_i}(-p)~\widetilde{\Xi_{\star_1}^i}(p)
\end{eqnarray}
\par \noindent the free part of the action. Moreover it can be seen that;
\begin{eqnarray} \label {56}
\widetilde{\Xi_{\star}^i}(p)=\widetilde{{\Xi}^{i~}}(p)~e^{\alpha (0,p)}~,
\end{eqnarray}
\par \noindent for $p\in\mathbb{R}^m$, $i=1,...,n$, and for translation-invariant star product $\star$ generated by $\alpha$. We note that $\widetilde{\Xi^\phi}$ is the Fourier transform of Laplacian operator for the standard Klein-Gordon filed $\phi$. One can assume this situation for simplicity. It can be seen that replacing $\phi$ by $\phi^{'}$ leads to
\begin{eqnarray} \label {57}
 S_{\star_1}^0=\sum_{i=1}^n\int \frac{\emph{\emph{d}}^m p}{(2\pi)^m}~\frac{1}{2} \tilde{\phi_i^{'}}(p)~\tilde{\phi_i^{'}}(-p)~\widetilde{\Xi_{\star_2}^i}(p)~.
\end{eqnarray}
\par For general kinds of quantum fields such as spinors and vectors, propagators and differential operators $\Xi$ may have two different indices, provided different fields are coupled in the free action, and consequently the factor of $\frac{1}{2}$ disappears from (\ref {55}) in such situations. However, in general for any given quantum field theory one finds that;
\begin{eqnarray} \label {58}
S_{\star_1}^0 (\phi_1,...,\phi_n)=S_{\star_2}^0 (\phi_1^{'},...,\phi_n^{'})~,
\end{eqnarray}
\par \noindent and hence for equivalent star products $\star_1$ and $\star_2$;
\begin{eqnarray} \label {59}
\widetilde{G_{\star_1}^{\phi_i,\phi_j}}=\widetilde{G_{\star_2}^{\phi_i^{'},\phi_j^{'}}}~,
\end{eqnarray}
\par \noindent where $\widetilde{G_{\star}^{\phi_i,\phi_j}}$ is the propagator of fields $\phi_i$ and $\phi_j$ , $i,j=1,...,n$, for star product $\star$. It is seen that for a fixed quantum field theory, two its different non-commutative copies with two $\alpha^*$-cohomologous complex translation-invariant star products $\star_1$ and $\star_2$, substantially define the same physics. More precisely, moving through an $\alpha^*$-cohomology class of $H_{\alpha^*}^2 (\mathbb{R}^m)$, produces no new physics. As a direct corollary the Wick-Voros non-commutative field theories, have no new quantum behaviors in compare with the Groenewold-Moyal ones. Therefore, all abnormal effects of Wick-Voros non-commutative field theories such as UV/IR mixing, non-locality and consequently non-renormalizability, coincide exactly with those of Groenewold-Moyal ones. Consequently, it would be naturally expected that the Grosse-Wulkenhaar approach \cite{GW1, GW2} and the method of $1/p^2$ \cite{GMRT} also work properly for renormalizing $\phi_{\star_{V-W}}^4$-theory. Moreover, for an interesting result we see that any translation-invariant version of $\phi^4$-theory (gauge theory) with a commutative star product is local, causal, unitary, and renormalizable. In fact, we have the following theorem;\\

\par \textbf{Theorem 5;} \emph{Any two translation-invariant (non-commutative) versions of a quantum field theory with equivalent star products coincide in all their quantum behaviors, such as for locality, causality, unitarity, renormalizability, UV/IR behaviors, the structures of Green's functions singularities, and consequently the scattering matrix.}\\


\par As it was stated above, the structures of 2- and 4-point functions of a translation-invariant $\phi_\star^4$-theory are described by $\omega$ as a character of the $\alpha^*$-cohomology class of its star product up to finite orders of loop calculations. In fact, as mentioned above for any given complex 2-cocycle $\alpha$, one finds; $\omega_\alpha(p,q)=-2\alpha_H(p,q)$, for $\alpha_H$ the harmonic form of $[\alpha]\in H_{\alpha^*}^2(\mathbb{R}^m)$. Thus;
\begin{eqnarray} \label {61}
\omega_\alpha(p,q)=2\alpha_H(q,p)~.
\end{eqnarray}
\par Strictly speaking, $\omega$ is essentially nothing more than the harmonic form and particularly gives no more information about $\alpha^*$-cohomology classes rather than harmonic elements.\\


\par It can be precisely shown that in all orders of loop calculations the quantum corrections for any translation-invariant quantum field theory are thoroughly described only by $\omega$ (the harmonic form) but no more characters of the $\alpha^*$-cohomology classes. To see this, at the first step it must be shown that the complex translation-invariant products described by coboundaries affect the Feynman diagrams amplitudes only in an exponential factor of external momenta. Indeed, it can be seen by induction that;
\begin{eqnarray} \label {63}
 \sum_{i=2}^n~\partial\beta(\sum_{j=1}^i~p^j,\sum_{j=1}^{i-1}~p^j)=\sum_{i=1}^n~\beta(p^i)-\beta(\sum_{i=1}^n~p^i)~,
\end{eqnarray}
\par \noindent for any $\beta\in \emph{\emph{C}}_*^1 (\mathbb{R}^m)$, and for any collection of $p^1,...,p^n\in \mathbb{R}^m$. Therefore, by momentum conservation law at vertices, the non-commutative vertex (\ref {51}) for complex translation-invariant product $\star$ generated by $\partial\beta$ is
\begin{eqnarray} \label {64}
V_\star^k (p^1,...,p^k )=e^{\sum_{i=1}^k\beta(p^i)}~.
\end{eqnarray}
\par On the other hand, the star product factor of propagators is;
\begin{eqnarray} \label {65}
e^{-\partial \beta(0,p)}=e^{-\beta(p)-\beta(-p)}~,
\end{eqnarray}
$p\in \mathbb{R}^m$. This cancels the relevant exponential factors of initial and final vertices. Therefore, (\ref {65}) together with (\ref {64}) cancels out all the internal momenta dependent factors and keeps only the exponential factors of external momenta.\\
\par Therefore, if $\alpha_1=\alpha_2+\partial\beta$, then the amplitudes of a fixed Feynman diagram for star products $\star_1$ and $\star_2$, respectively generated by $\alpha_1$ and $\alpha_2$, coincide up to a factor of external momenta. This achievement lets one to study the role of complex translation-invariant star products in loop calculations only for harmonic forms. On the other hand, since $\alpha(0,p)=0$, $p\in \mathbb{R}^m$, for harmonic form $\alpha$, then there is no nontrivial star product factor for the propagators. Moreover, by (\ref {51}) and (\ref {61}) the non-commutative vertex is;
\begin{eqnarray} \label {66}
V_\star^k (p^1,…,p^k )=e^{\sum_{i=1}^{k-1}~\frac{1}{2}\omega_\alpha(\sum_{j=1}^ip^j,p^{i+1})}~,
\end{eqnarray}
 \par \noindent for $\star$ a translation-invariant star product generated by $\alpha$. This shows that the quantum corrections of translation-invariant quantum field theories not only are described thoroughly by the $\alpha^*$-cohomology classes of star products, but they are precisely explained by $\omega$ or the harmonic forms of $\alpha^*$-cohomology classes due to \emph{the Hodge theorem}. Therefore, we conclude that;\\

\par \textbf{Theorem 6;} \emph{The regularization methods of a translation-invariant (non-commutative) quantum field theory work properly for all its other translation-invariant (non-commutative) versions with equivalent star products. Moreover, such regularization methods are thoroughly reflected by the generator of equivalent harmonic star product.}


\par
\section{The Origin of $\alpha^*$-Cohomology and the Quantum Equivalence Theorem}
\setcounter{equation} {0}
\par

More than the algebraic approach of $\alpha^*$-cohomology as a classifying program, there can be defined other classification methods for translation-invariant quantum field theories. From the viewpoints of quantum physics the most primitive equivalence relation for two deformation quantization star products $\star_1$ and $\star_2$ is the equality under integration. We refer to this concept by equivalence under integration. More precisely, two star products $\star_1$ and $\star_2$ are called to be n-equivalent under integration, if and only if for any $1\leq k\leq n$ and for any set of $f_1,...,f_k\in \mathcal{S}_c (\mathbb{R}^m )$;
\begin{eqnarray} \label {67}
\int_{\mathbb{R}^m} ~f_1 \star_1...\star_1 f_k =\int_{\mathbb{R}^m} ~f_1 \star_2...\star_2 f_k~.
\end{eqnarray}
\par But it can be easily seen that in the case of translation-invariant star products, if $\star_1$ and $\star_2$ are n-equivalent under integration for $n\geq3$, then $\star_1=\star_2$. Thus, since the interaction term in Lagrangian densities includes at least 3 quantum fields, then, from Lagrangian formalism points of view the equivalence under integration leads to a trivial classification of translation-invariant quantum field theories. Therefore, to have a more general classification, one should appropriately weaken the equivalence relation of (\ref {67}). The most general idea is to set two translation-invariant star products $\star_1$ and $\star_2$ in an equivalence class if and only if they lead to the same scattering matrix for any given (renormalizable) quantum field theory. We refer to this meaning of classification by quantum equivalence. Indeed, due to the LSZ formula for scattering matrix, if two translation-invariant star products $\star_1$ and $\star_2$ are quantum equivalent, then for any fixed given quantum field theory including fields $\{\phi_i\}_{i=1}^k$, two translation-invariant non-commutative versions with star products $\star_1$ and $\star_2$, lead to the following equalities for all connected n-point functions;
\begin{eqnarray} \label {68}
\langle\tilde{\phi_{i_1}}(p_1 )...\tilde{\phi_{i_n}}(p_n)\rangle_{\star_1}=e^{\beta_{i_1}(p_1)}...e^{\beta_{i_n}(p_n )}~\langle\tilde{\phi_{i_1}}(p_1 )...\tilde{\phi_{i_n}}(p_n)\rangle_{\star_2}~,
\end{eqnarray}
\par \noindent $p_1,...,p_n\in \mathbb{R}^m$, for all $n\geq1$ and for field dependent smooth functions $\{\beta_i\}_{i=1}^k$. But, since the types of quantum fields are unimportant for star products, one should set; $\beta_1=...=\beta_k=\beta$. Therefore, (\ref {68}) can be written in the following form;
\begin{eqnarray} \label {69}
\frac{\int_{\phi_{i_1}...\phi_{i_k}}~e^{iS_{\star_1}}~\tilde{\phi_{i_1}}(p_1)~...~\tilde{\phi_{i_k}}(p_n)}{\int_{\phi_{i_1}...\phi_{i_k}}~e^{iS_{\star_1}}}
=e^{\sum_{i=1}^n\beta(p_i)}~\frac{\int_{\phi_{i_1}...\phi_{i_k}}~e^{iS_{\star_2}}~\tilde{\phi_{i_1}}(p_1)~...~\tilde{\phi_{i_k}}(p_n)}{\int_{\phi_{i_1}...\phi_{i_k}}~e^{iS_{\star_2}}}~.
\end{eqnarray}
\par By using the transformation $\tilde{\phi_i}(p)\rightarrow e^{\beta(p)}~\tilde{\phi_i}(p)$ in the left hand side of (\ref {69}), it would be clear that this equality holds if and only if for any $n\geq1$, and for any set of $f_1,...,f_n\in \mathcal{S}_c(\mathbb{R}^m)$, one finds;
\begin{eqnarray} \label {70}
\int_{\mathbb{R}^m}~f_1^{'}\star_1...\star_1f_n^{'}=\int_{\mathbb{R}^m}~f_1\star_2...\star_2f_n~,
\end{eqnarray}
\par \noindent where;
\begin{eqnarray} \label {71}
f^{'}(x)=\int \frac{\emph{\emph{d}}^m p}{(2\pi)^m}~e^{ip.x}~ \tilde{f}(p)~e^{\beta(p)}~,
\end{eqnarray}
\par \noindent $f\in S_{c,1}(\mathbb{R}^m)$. By this, (\ref {70}) can be considered as the most appropriate general form for (\ref {67}).\\
 \par Quantum equivalence, i.e. equation (\ref {70}), should be studied step by step. For $n=1,2,3$, (\ref {70}) respectively leads to;
\begin{eqnarray} \label {72}
\beta(0)=0~,
\end{eqnarray}

\begin{eqnarray} \label {73}
\alpha_1(0,p)+\beta(-p)+\beta(p)=\alpha_2(0,p)~,
\end{eqnarray}
\par \noindent $p\in \mathbb{R}^m$, and;
\begin{eqnarray} \label {74}
\alpha_1(0,-p-q)+\alpha_1(p+q,p)+\beta(p)+\beta(q)+\beta(-p-q)
=\alpha_2(0,-p-q)+\alpha_2(p+q,p)~,
\end{eqnarray}
\par \noindent $p,q\in \mathbb{R}^m$ for $\alpha_1$ and $\alpha_2$ respectively the generators of $\star_1$ and $\star_2$. Equation (\ref {72}) asserts that; $\beta\in \emph{\emph{C}}^1 (R^m)$, while combining (\ref {73}) and (\ref {74}) results in:
\begin{eqnarray} \label {75}
\alpha_1+\partial\beta=\alpha_2~.
\end{eqnarray}
 \par On the other hand, since $\star_1$ and $\star_2$ are associative, then; $\partial\alpha_1=\partial\alpha_2=0$. It is now obvious that quantum equivalence leads naturally to $\alpha$-cohomology. Moreover, if one imposes the complexity structure (\ref {41}), quantum equivalence would strictly be considered as the original root for $\alpha^*$-cohomology. In fact, it has been proven that translation-invariant star products $\star_1$ and $\star_2$, respectively generated by 2-cocycles $\alpha_1$ and $\alpha_2$, lead to the same physics if $\alpha_1$ and $\alpha_2$ are $\alpha^*$-cohomologous. Actually this fact is the inverse to what we stated in \emph{theorem 5}. Therefore, we have just shown;\\

\par \textbf{Theorem 7;} \emph{Two complex translation-invariant star products $\star_1$ and $\star_2$ are equivalent if and only if there exists a fixed $\beta\in C^\infty(\mathbb{R}^m)$, with $\beta(0)=0$, such that for any $n\geq1$, the equality}
\begin{eqnarray} \label {th}
\widetilde{G_{\star_1~conn.}} (p_1,...,p_n)=e^{\sum_{i=1}^n\beta(p_i)}~\widetilde{G_{\star_2~conn.}} (p_1,...,p_n)
\end{eqnarray}
\par \noindent \emph{holds for any given (renormalizable) quantum field theory, where $G_{conn.}$ is any connected $n$-point function, $G_{\star~conn.}$ is its non-commutative version for the star product $\star$ and $\widetilde{G_{\star~conn.}}(p_1,...,p_n)$ is its Fourier transform for the modes $\{p_i\}_{i=1}^n$. Therefore, $\alpha^*$-cohomology produces the most general setting for classification of translation-invariant quantum field theories via the view points of quantum physics.}\\

\par From now on this theorem is referred to as \emph{the quantum equivalence theorem}. Using the LSZ formula the second version of \emph{the quantum equivalence theorem} will be trivial:\\

\par \textbf{Theorem 8;} \emph{Two complex translation-invariant star products $\star_1$ and $\star_2$ are equivalent if and only if for any given (renormalizable) quantum field theory, its two translation-invariant (non-commutative) versions with $\star_1$ and $\star_2$ have the same scattering matrix.}\\


\par
\section{Algebraic Structures of Translation-Invariant Products}
\setcounter{equation}{0}

\par

Using (\ref {48}) and the properties of harmonic forms in (\ref {33}) one finds that;
\begin{eqnarray} \label {76}
\left\{
  \begin{array}{ll}
    {\omega(p,q)=-\omega(q,p)} \\
    {\omega(p,q)=\omega(-p,-q)~,} \\
    {\omega(p,0)=0}
  \end{array}
\right.
\end{eqnarray}
\par \noindent for any $p,q\in \mathbb{R}^m$.\\
 \par On the other hand, (\ref {7}) asserts that (\ref {48}) can be written in the form of;
\begin{eqnarray} \label {77}
\omega_\alpha(p,q)=\alpha(p,p-q)-\alpha(p,q)=\omega_\alpha(p-q,q)~,
\end{eqnarray}
\par \noindent and consequently;
\begin{eqnarray} \label {78}
\omega(p,q)=\omega(p+nq,q)
\end{eqnarray}
\par \noindent for any $n\in \mathbb{Z}$. The property (\ref {78}) together with (\ref {61}), proves (\ref {30}) by (\ref {48}). Thus, $\omega$ satisfies the associativity condition of (\ref {7}). This consequently confirms the cohomological description of loop calculations in translation-invariant field theories. On the other hand, again by uniqueness of harmonic elements, (\ref {78}) and (\ref {61}) lead to (\ref {31}). Moreover, by (\ref {76}) and (\ref {78}) it is seen that;
\begin{eqnarray} \label {80}
\omega(p,q)=-\omega(p,-q)~,
\end{eqnarray}
\par \noindent and thus; $\alpha_H(p,q)=-\alpha_H (p,-q)$.\\
 \par In appendix B it is shown that, (\ref {78}) leads to
\begin{eqnarray} \label {81}
\omega(rp,p)=0~,
\end{eqnarray}
\par \noindent for any $r\in \mathbb{Q}$, and thus by continuity of $\omega$;
\begin{eqnarray} \label {82}
\omega(rp,p)=0~,
\end{eqnarray}
\par \noindent for any  $r\in \mathbb{R}$.\\
 \par It should be noted that the proof of equalities (\ref {78}), (\ref {80}) and (\ref {82}) need no use of complexity condition due to (\ref {41}). Therefore, all these equalities hold for any general harmonic form of $H_\alpha^2(\mathbb{R}^m)$, by replacing $\omega$ with $\alpha_H$.\\
\par What we are looking for is a simple criteria to distinguish two equivalent star products. To answer this question, a natural approach is to extend the domain of star products to the algebra of polynomials generated by $\{x^\mu\}_{\mu=1}^m$. This leads to non-commutative structures of space-time manifold. More precisely, the non-commutative structure of space-time due to star product $\star$ generated by $\alpha$ is given by;
\begin{eqnarray} \label {var1}
[x^\mu,x^\nu]_\star =x^\mu\star x^\nu-x^\mu\star x^\nu=\frac{\partial^2 \alpha}{\partial z_\nu^1 \partial z_\mu^2}(0,0)
-\frac{\partial^2\alpha}{\partial z_\mu^1 \partial z_\nu^2}(0,0)~,
\end{eqnarray}
\par \noindent for the coordinate functions $\{z_\mu^1, z_\mu^2\}_{\mu=1}^m$ for $\mathbb{R}^m\times\mathbb{R}^m$ dual to $\{x^\mu\}_{\mu=1}^m$ due to the Fourier transform. It can be seen that if $\star$ is commutative, i.e.; $\alpha=\partial\beta$, for 1-cochain $\beta$, then;
\begin{eqnarray} \label {var2}
[x^\mu,x^\nu]_\star=0~,
\end{eqnarray}
\par \noindent $\mu,\nu=1,...,m$. Equality (\ref {var2}) shows that the non-commutative structure of space-time is particularly given by the $\alpha$-cohomology class of the 2-cocycle.  Actually if $\star_1\thicksim\star_2$ then;
\begin{eqnarray} \label {var3}
[x^\mu,x^\nu]_{\star_1}=[x^\mu,x^\nu]_{\star_2}~,
\end{eqnarray}
\par \noindent $\mu,\nu=1,...,m$. This leads to a well-defined map which corresponds to any $\alpha$-cohomology class a non-commutative structure of space-time. Although using different forms of Groenewold-Moyal star products clears the surjectivity of this map, but its injectivity is not clear at all (see \cite{Varshovi2}). Therefore, to give a more effective criteria for equivalent translation-invariant star products the domain of star products should be extended to Fourier modes, and then, the commutation relations of Fourier modes for such star products should be studied. Consider a 2-cocycle $\alpha$ and its star product $\star$. It is easy to see that;
\begin{eqnarray} \label {83}
[e^{ip.x},e^{iq.x}]_\star
=(e^{\alpha_H(p+q,p)}-e^{-\alpha_H(p+q,p)})~e^{\partial\beta(p+q,p)}~e^{i(p+q).x}~,
\end{eqnarray}
\par \noindent $p,q\in \mathbb{R}^m$, with $\alpha=\alpha_H+\partial\beta$ due to \emph{the Hodge theorem}. Thus, for two $\alpha$-cohomologous 2-cocycles $\alpha_1$ and $\alpha_2$ with $\alpha_1=\alpha_2+\partial\beta$, and for any $p,q\in \mathbb{R}^m$, one finds;
\begin{eqnarray} \label {85}
[e^{ip.x},e^{iq.x}]_{\star_1}=e^{\partial\beta(p+q,p)}~[e^{ip.x},e^{iq.x}]_{\star_2}~.
\end{eqnarray}
 \par Therefore, $\alpha$-cohomology classes reveal the non-commutative structure of space-time up to an exponential factor for a coboundary term. Conversely, it can be seen that if star products $\star_1$ and $\star_2$ satisfy condition (\ref {85}) for all Fourier modes and for a smooth function $\beta$ with $\beta(0)=0$, then $\star_1\sim\star_2$. To see this fact, consider two translation-invariant star products $\star_1$ and $\star_2$ which satisfy (\ref {85}) for some 1-cochain $\beta$. Therefore;
\begin{eqnarray} \label {87}
(e^{\alpha_H^1(p+q,p)}-e^{-\alpha_H^1(p+q,p)})=(e^{\alpha_H^2(p+q,p)}-e^{-\alpha_H^2(p+q,p)})~e^{\partial\gamma(p+q,p)}~,
\end{eqnarray}
\par \noindent $p,q\in \mathbb{R}^m$, for $\gamma=\beta+\beta^2-\beta^1$, where $\alpha_i=\alpha_H^i+\partial\beta^i$, is the generator of $\star_i$, $i=1,2$. \\

\par By (\ref {33}), $e^{\alpha_H^i(p+q,p)}-e^{-\alpha_H^i(p+q,p)}$, $i=1,2$, are anti-symmetric for exchange of $p\rightleftharpoons p+q$. Thus; $\partial\gamma(p+q,p)=\partial\gamma(p,p+q)$, for any $p,q\in \mathbb{R}^m$. Setting $p=0$, one finds that $\gamma(q)=-\gamma(-q)$, for any $q\in \mathbb{R}^m$. Therefore, we have; $\partial\gamma=0$, which by (\ref {87}) and (\ref {31}) leads to;
\begin{eqnarray} \label {91}
e^{\alpha_H^1(p,q)}-e^{-\alpha_H^1(p,q)}=e^{\alpha_H^2(p,q)}-e^{-\alpha_H^2(p,q)}~,
\end{eqnarray}
\par \noindent for any $p,q\in \mathbb{R}^m$. Then, one has either $e^{\alpha_H^1(p,q)}=e^{\alpha_H^2(p,q)}$ or $e^{\alpha_H^1(p,q)+\alpha_H^2(p,q)}=-1$. The former is equivalent to $\alpha_H^1=\alpha_H^2$ near the origin, but the later can also be satisfied when $\alpha_H^1\neq\alpha_H^2$. In appendix C, it is shown that $\alpha_H^1\neq\alpha_H^2$ and (\ref {82}) lead to a contradiction. Thus, one finds that $\alpha_H^1=\alpha_H^2$ and consequently; $\star_1\sim\star_2$. This proves that;\\

\par \textbf{Theorem 9;} \emph{Two translation-invariant star products $\star_1$ and $\star_2$ are equivalent if and only if there exists a fixed 1-cochain $\beta$ such that the equality (\ref {85}) holds for any $p,q\in \mathbb{R}^m$.}\\


\section{Conclusions}
\par In this article, translation-invariant star product was discussed in the setting of $\alpha$-cohomology. It was explicitly shown that the second $\alpha$-cohomology group classifies translation-invariant star products up to commutative products. Then $\alpha^*$-cohomology as a sub-theory of $\alpha$-cohomology defined to classify complex translation-invariant star products. Moreover, an algebraic version of Hodge theorem was derived for the second $\alpha$-cohomology group which led to the definition of harmonic translation-invariant products, the unique elements of $\alpha$-cohomology classes with some special properties. It was also shown that the loop calculation in any translation-invariant non-commutative quantum field theory is thoroughly described by the $\alpha^*$-cohomology class of its star product. In fact, it was seen that moving through an $\alpha^*$-cohomology class produces no new physics. This showed that the harmonic translation-invariant products due to the Hodge theorem, play a crucial role in studding the physics of translation-invariant quantum field theories. Conversely, it was precisely shown that the inverse conclusion is also true, i.e.; if two complex translation-invariant star products $\star_1$ and $\star_2$ lead to the same physics for any given quantum field theory, then their relevant generators are $\alpha^*$-cohomologous. Finally, it was discussed that $\alpha^*$-cohomology is essentially the most general classification for translation-invariant versions of quantum field theories. These results guided us to the quantum equivalence theorem which asserts that two star products $\star_1$ and $\star_2$ are equivalent if and only if for any given (renormalizable) quantum field theory they lead to the same scattering matrix when are replaced with the ordinary product.\\


\section{Acknowledgments}
\par My special thanks go to Ahmad Shafiei Deh Abad for his ever warm welcome, patience and useful comments in geometric viewpoints and intuitions. It would be my honor to dedicate this article to Ahmad Shafiei Deh Abad for his 68th birth day on Day 5th. Also I have to confess with regards that the author owes most of this article to special considerations of M. M. Sheikh-Jabbari, his elegant ideas and his fruit-full comments. On the other hand, I should say my kind gratitude to M. Amini for hints and discussions. Moreover, it is important to say that almost the whole of this work was made during the PhD period of the author at the Sharif University of Technology, under the supervision of F. Ardalan. Finally, my deepest regards go to S. Ziaee for all her kindnesses.

\par
\section{Appendices}

\begin{appendix}\setcounter{equation}{0}
\section{ }
\par

In this appendix, it is shown that $\alpha\in\emph{\emph{C}}^2(\mathbb{R}^m)$ is a coboundary if;
\begin{eqnarray} \label {A1}
\left\{
  \begin{array}{ll}
    {\partial\alpha=0} \\
    {\alpha(p,q)=\alpha(p,p-q)} \\
    {\alpha(p,q)=-\alpha(-p,-q)} \\
    {\alpha(p,q)=-\alpha(q,p)}
  \end{array}
\right.
~.
\end{eqnarray}


\par \noindent for any $p,q\in \mathbb{R}^m$. To this end initially, choose the coordinate system of;
\begin{eqnarray} \label {A2}
\left\{
  \begin{array}{ll}
    {w=p-q} \\
    {z=q}
  \end{array}
\right.
~.
\end{eqnarray}\\


\par It can be shown that;
\begin{eqnarray} \label {A3}
 \frac{\partial^2\alpha}{\partial z^\mu \partial w^\nu}(p,q)=\frac{\partial^2\alpha}{\partial q^\mu\partial p^\nu}(p,0)
\end{eqnarray}
\par \noindent for any $p,q\in \mathbb{R}^m$ and for any $1\leq \mu,\nu\leq m$. Indeed;
\begin{eqnarray} \label {A4}
 \frac{\partial^2\alpha}{\partial z^\mu \partial w^\nu}(p,q)=\frac{\partial^2\alpha}{\partial p^\mu \partial p^\nu}(p,q)+\frac{\partial^2\alpha}{\partial q^\mu \partial p^\nu}(p,q)
  =\frac{\emph{\emph{d}}^2}{\emph{\emph{d}}r\emph{\emph{d}}s}|_{r=s=0} ~\alpha(p+re^\nu+se^\mu,q+se^\mu)~,
\end{eqnarray}
\par \noindent for $\{e^\mu\}?_{\mu=1}^m$, the standard basis of $\mathbb{R}^m$ dual to coordinate functions $\{x^\mu\}_{\mu=1}^m$ due to the Fourier transform. Using (\ref {7}) one finds that;
\begin{eqnarray} \label {A5}
  \frac{\emph{\emph{d}}^2}{\emph{\emph{d}}r\emph{\emph{d}}s}\alpha(p+re^\nu+se^\mu,q+se^\mu)
  =\frac{\emph{\emph{d}}^2}{\emph{\emph{d}}r\emph{\emph{d}}s}(\alpha(p+re^\nu,q)-\alpha(p+re^\nu,-se^\mu)+\alpha(q,-se^\mu))~,
\end{eqnarray}
\par \noindent and then (\ref {A3}) follows.\\

 \par Therefore;
\begin{eqnarray} \label {A6}
 \frac{\partial}{\partial q^\sigma} \frac{\partial^2}{\partial z^\mu \partial w^\nu}\alpha=0
\end{eqnarray}
\par \noindent for any $1\leq \mu,\nu,\sigma\leq m$. In the other words, one finds that;
\begin{eqnarray} \label {A7}
 (\frac{\partial}{\partial z^\sigma}-\frac{\partial}{\partial w^\sigma})\frac{\partial^2}{\partial z^\mu \partial w^\nu}~\alpha=0~.
\end{eqnarray}
 \par Now suppose that;
\begin{eqnarray} \label {A8}
\alpha(w,z)=f_1(w)+f_2(w,z)+f_3(z)
\end{eqnarray}
\par \noindent for $f_1,f_2,f_3\in \emph{\emph{C}}^\infty (\mathbb{R}^m)$ such that;
\begin{eqnarray} \label {A9}
f_2 (w,z)\neq g_1 (w)+g_2 (w,z)~,
f_2 (w,z)\neq h_1 (z)+h_2 (w,z)~,
\end{eqnarray}
\par \noindent $g_1,g_2,h_1,h_2\in \emph{\emph{C}}^\infty (\mathbb{R}^m)$, for non-constant $g_1$ and $h_1$. Thus (\ref {A7}) leads to;
\begin{eqnarray} \label {A10}
 (\frac{\partial}{\partial z^\sigma}-\frac{\partial}{\partial w^k}) \frac{\partial^2}{\partial z^\mu \partial w^\nu}~f_2=0~.
\end{eqnarray}


 \par Now choose the coordinate system of
\begin{eqnarray} \label {A11}
\left\{
  \begin{array}{ll}
    {\eta=z+w} \\
    {\xi=z-w}
  \end{array}
\right.
~.
\end{eqnarray}
 \par Therefore, by (\ref {A10}) and (\ref {A11}) one finds that;
\begin{eqnarray} \label {A12}
 \frac{\partial}{\partial\xi^k} \frac{\partial^2}{\partial z^\mu \partial w^\nu}~f_2=0~.
\end{eqnarray}
\par Consequently, we have;
\begin{eqnarray} \label {A13}
 \frac{\partial^2}{\partial z^\mu \partial w^\nu}f_2(\eta,\xi)=h_{\mu\nu} (\eta)~.
\end{eqnarray}
 \par Then since
\begin{eqnarray} \label {A14}
 \frac{\partial^2}{\partial z^\mu \partial w^\nu}=\frac{\partial ^2}{\partial\eta^\mu \partial\eta^\nu}-\frac{\partial^2}{\partial\eta^\mu \partial\xi^\nu}+\frac{\partial^2}{\partial\xi^\mu \partial\eta^\nu}-\frac{\partial^2}{\partial\xi^\mu \partial\xi^\nu}~,
\end{eqnarray}
\par \noindent (\ref {A13}) lets one to set; $f_2 (\eta,\xi)=g(\eta)+\xi^\mu g_\mu (\eta)+\xi^\mu \xi^\nu g_{\mu\nu} (\eta)$ , for some $g,g_\mu,g_{\mu\nu}\in \emph{\emph{C}}^\infty (\mathbb{R}^m)$, $1\leq\mu,\nu\leq m$. Therefore;
\begin{eqnarray} \label {A15}
\alpha(w,z)=f_1 (w)+g(z+w)+(z-w)^\mu g_\mu (z+w)
+(z-w)^\mu (z-w)^\nu g_{\mu\nu} (z+w)+f_3 (z)~.
\end{eqnarray}


 \par Note that for the second property of (\ref {A1}) (the commutativity condition), then; $\alpha(w,z)=\alpha(z,w)$. This implies that;
\begin{eqnarray} \label {A16}
(f_3 (w)-f_1 (w))-(f_3 (z)-f_1 (z))=2(z-w)^\mu g_\mu (z+w)~.
\end{eqnarray}
 \par Then, (ref {A9}) leads to;
\begin{eqnarray} \label {A17}
g_\mu=0~~~,~~~f_1=f_3=f~,
\end{eqnarray}
 \par \noindent $\mu=1,...,m$. Therefore;
\begin{eqnarray} \label {A18}
\alpha(p,q)=f(q)+g(p)+(2q-p)^\mu (2q-p)^\nu g_{\mu\nu} (p)+f(p-q)~,
\end{eqnarray}
\par \noindent for any $p,q\in \mathbb{R}^m$. Now by $\alpha(-p,-q)=\alpha(q,p)$ one finds that;
\begin{eqnarray} \label {A19}
f(-q)+g(-p)+(2q-p)\mu (2q-p)^\nu g_{\mu\nu} (-p)
=f(p)+g(q)+(2p-q)^\mu (2p-q)^\nu g_{\mu\nu} (q)~.
\end{eqnarray}


 \par Acting $\partial^2/\partial p^\mu \partial q^\nu$ on both sides of (\ref {A19}) yields the following result:
\begin{eqnarray} \label {A20}
 4g_{\mu\nu}(-p)+4(2q-p)^\sigma\frac{\partial}{\partial p^\mu }g_{\nu\sigma}(-p)=4g_{\mu\nu}(q)-4(2p-q)^\sigma\frac{\partial}{\partial q^\nu}g_{\mu\sigma}(q)~.
\end{eqnarray}
 \par Thus, one finds that;
\begin{eqnarray} \label {A21}
 p^\sigma\frac{\partial}{\partial p^\mu}g_{\nu\sigma}(p)=-p^\sigma\frac{\partial}{\partial p^\nu}g_{\mu\sigma}(p)~.
\end{eqnarray}
 \par Acting $\partial/\partial q^\sigma$ on both sides of (\ref {A20}) and using (\ref {A21}) one also finds that;
\begin{eqnarray} \label {A22}
 p^\sigma\frac{\partial}{\partial p^\mu}g_{\nu\sigma}(-p)=\frac{1}{2}p^\sigma(2p-q)^\lambda\frac{\partial^2}{\partial q^\sigma\partial q^\nu}g_{\mu\lambda}(q)~.
\end{eqnarray}
\par Setting $q=2p$, gives rise to;
\begin{eqnarray} \label {A23}
 p^\mu\frac{\partial}{\partial p^\mu}(g_{\nu\sigma}(p)p^\sigma)=g_{\nu\sigma}(p)p^\sigma~.
\end{eqnarray}


 \par It can be seen that the solution of differential equation (\ref {A23}) results in:
\begin{eqnarray} \label {A24}
g_{\mu\nu}(p)=c_{\mu\nu}\in \mathbb{C}~,
\end{eqnarray}
\par \noindent $\mu,\nu=1,...,m$. Thus;
\begin{eqnarray} \label {A25}
f_2(w,z)=g(w+z)-2w^\mu z^\nu c_{\mu\nu}+w^{\mu}w^{\nu} c_{\mu\nu}+z^\mu z^\nu c_{\mu\nu}~,
\end{eqnarray}
\par \noindent which according to (\ref {A9}) yields $c_{\mu\nu}=0$ for any $\mu$ and $\nu$. Therefore;
\begin{eqnarray} \label {A26}
f_2 (w,z)=g(w+z)~.
\end{eqnarray}
\par Consequently, by (\ref {A19}) it follows that;
\begin{eqnarray} \label {A27}
f(-q)-g(q)=f(p)-g(-p)~,
\end{eqnarray}
\par \noindent which results in;
\begin{eqnarray} \label {A28}
g(p)=f(-p)+c_0~,
\end{eqnarray}
\par \noindent for any $p\in \mathbb{R}^m$ and for a complex number $c_0$. Thus, according to (\ref {A18}) we have;
\begin{eqnarray} \label {A29}
\alpha(p,q)=f(q)+f(-p)+f(p-q)+c_0
\end{eqnarray}
\par \noindent for any $p,q\in \mathbb{R}^m$. But (\ref {8}) and the third equality of (\ref {A1}) lead to;
\begin{eqnarray} \label {A30}
f(-p)=-f(p)+c_1~,
\end{eqnarray}
\par \noindent for any $p\in \mathbb{R}^m$ and for $c_1=-f(0)-c_0$. Therefore, (\ref {A28}) takes the following form;
\begin{eqnarray} \label {A31}
\alpha(p,q)=f(q)-f(p)+f(p-q)+c_1~.
\end{eqnarray}
\par Finally the fourth equality of (\ref {A1}), implies that $c_1=0$ and eventually
\begin{eqnarray} \label {A32}
\alpha(p,q)=f(q)-f(p)+f(p-q)~,
\end{eqnarray}
\par \noindent for any $p,q\in \mathbb{R}^m$. On the other hand, by (\ref {8}) we have; $f(0)=0$. Therefore, $f\in \emph{\emph{C}}^1 (\mathbb{R}^m)$ and consequently according to (\ref {A32}); $\alpha=\partial f$.\\

\par
\section{}
\par

In this appendix, it is shown that (\ref {78}) concretely leads to (\ref {81}). To see this note that by (\ref {78});

\begin{eqnarray} \label {B1}
\omega(p,p)=\omega(np,p)=0~,
\end{eqnarray}
\par \noindent for any $n\in \mathbb{Z}$. Then, the iterated form of (\ref {B1}) due to (\ref {78}) would be
\begin{eqnarray} \label {B2}
\omega(N_k p,N_{k-1} p)=0~,
\end{eqnarray}
\par \noindent $k\geq1$, for recursive formulae;
\begin{eqnarray} \label {B3}
N_k=n_k N_{k-1}+N_{k-2}~,
\end{eqnarray}
\par \noindent with $n_k\in \mathbb{Z}$, $k\geq2$, $N_0=1$ and $N_1=n_1\in \mathbb{Z}$. Clearly (\ref {B2}) can be rewritten in the form of \begin{eqnarray} \label {B4}
\omega(\frac{N_k}{N_{k-1}}p,p)=0~.
\end{eqnarray}
\par But according to (\ref {B3}) it can be seen that;
\begin{eqnarray} \label {B5}
\frac{N_k}{N_{k-1}}=n_k+\frac{1}{n_{k-1}+\frac{1}{n_{k-2}+\frac{1}{\ddots n_2+\frac{1}{n_1}}}}~.
\end{eqnarray}
 \par On the other hand, it is known \cite{HW} that for any rational number $r\in \mathbb{Q}$, there is a finite sequence of integers $n_i$, $1\leq i\leq k$, with $n_k\in \mathbb{Z}$ and $n_i>0$ for $1\leq i<k$, such that
\begin{eqnarray} \label {B6}
r=\frac{N_k}{N_{k-1}}
\end{eqnarray}
\par \noindent in accordance to (\ref {B5}). This together with (\ref {B4}) proves (\ref {81}).\\
\par As mentioned in section $\emph{\emph{VIII}}$, the continuity of $\omega$ consequently leads to (\ref {82}). On the other hand, (\ref {82}) shows that any translation-invariant product on $\mathcal{S}_{c,1}(\mathbb{R})$, is $\alpha$-cohomologous to the ordinary point-wise product and thus is commutative. In the other words, (\ref {82}) results in $H_\alpha^2 (\mathbb{R})=0$, which shows that there is no translation-invariant non-commutative deformation quantization on $C^\infty (\mathbb{R})$.\\


\par
\section{}
\par
In this appendix, it is shown that (\ref {91}) and (\ref {82}) lead to $\alpha_H^1=\alpha_H^2$. As it was stated above (\ref {91}) leads to either $e^{\alpha_H^1(p,q)}=e^{\alpha_H^2(p,q)}$ or $e^{\alpha_H^1(p,q)+\alpha_H^2(p,q)}=-1$, $p,q\in \mathbb{R}^m$. It is claimed that $\alpha_H^1$ and $\alpha_H^2$ coincide everywhere. To see this fact precisely, suppose that $\alpha_H^1\neq \alpha_H^2$ over open set $V\subseteq \mathbb{R}^m\times \mathbb{R}^m$. Thus, $V$ breaks down into two possibly intersecting subsets; $V=V_1\cup V_2$, defined by: For any $(p,q)\in V_1$; $e^{\alpha_H^1(p,q)}=e^{\alpha_H^2(p,q)}$, but:
\begin{eqnarray} \label {C1}
Re(\alpha_H^1 )=Re(\alpha_H^2 )~~~~and~~~~Im(\alpha_H^1 )=Im(\alpha_H^2)+2k\pi~,
\end{eqnarray}
\par \noindent for some non-zero fixed $k\in \mathbb{Z}$ over each connected component of $V_1\subseteq V$. On the other hand, for any $(p,q)\in V_2$; $e^{\alpha_H^1(p,q)} e^{\alpha_H^2(p,q)}=-1$, or;
\begin{eqnarray} \label {C2}
Re(\alpha_H^1 )=-Re(\alpha_H^2)~~~~and~~~~Im(\alpha_H^1)+Im(\alpha_H^2)=(2k+1)\pi~,
\end{eqnarray}
\par \noindent for some fixed $k\in \mathbb{Z}$ over each connected component of $V_2$.\\
\par Now, suppose that the interior of $V_2\setminus V_1$ is not empty. Fix $(p,q)\in int(V_2\setminus V_1)$ and choose $\epsilon>0$ such that; $B_\epsilon(p,q)\subset int(V_2\backslash V_1)$, where  $B_\epsilon(p,q)$ is a ball with center $(p,q)$ and radius $\epsilon$. Thus for any $r\in \mathbb{R}^m$, with $|r|<\epsilon$, (\ref {7}) leads to;
\begin{eqnarray} \label {C3}
Im(\alpha(p,r))=Im(\alpha(q,r))~,
\end{eqnarray}
\par \noindent for $\alpha=\alpha_H^1+\alpha_H^2$. Choose $N\in \mathbb{N}$ such that $|q|<N\epsilon$. Therefore;
\begin{eqnarray} \label {C4}
Im(\alpha(p,q/N))=0~,
\end{eqnarray}
\par \noindent since $\alpha(q,q/N)=0$ due to (\ref {82}). Now by (\ref {7}), (\ref {C3}) and (\ref {C4}) it is seen that;
\begin{eqnarray} \label {C5}
Im(\alpha(p-nq/N,q-(n+1)q/N))=Im(\alpha(p-(n+1)q/N,q-(n+2)q/N))~,
\end{eqnarray}
\par \noindent for any $n\geq0$, which leads to the following equalities;
\begin{eqnarray} \label {C6}
\begin{array}{c}
  Im(\alpha(p,q))=Im(\alpha(p,q-q/N))=Im(\alpha(p-q/N,q-2q/N))~~~~~~~~~~~~ \\
  =...=Im(\alpha(p-nq/N,q-(n+1)q/N))=...~,
\end{array}
\end{eqnarray}
\par \noindent where the first equality is deduced from $B_\epsilon(p,q)\subset int(V_2\backslash V_1)$ and (\ref {C2}). It is now sufficient to set $n=N-1$ in (\ref {C6}) and find; $Im(\alpha(p,q))=0$, which leads to a contradiction with (\ref {C2}). Therefore, $int(V_2\backslash V_1 )=\emptyset$ and consequently; $V=V_1$, since $\alpha_H^1$ and $\alpha_H^2$ both are continuous. Thus, $e^{\alpha_H^1(p,q)}=e^{\alpha_H^2(p,q)}$ for any $p,q\in\mathbb{R}^m$ and therefore; $\alpha_H^1=\alpha_H^2+2ik\pi$, for a fixed $k\in \mathbb{Z}$ over $\mathbb{R}^m\times \mathbb{R}^m$. But since $\alpha_H^1$ and $\alpha_H^2$ coincide at the origin, $V=\emptyset$ and hence; $\alpha_H^1=\alpha_H^2$ everywhere on $\mathbb{R}^m\times \mathbb{R}^m$.

\end{appendix}



\begin{thebibliography}{99}
\bibitem{CM}
	A. Connes and M. Marcolli, \emph{Noncommutative Geometry, Quantum Fields and Motives}, American Mathematical Society, Hindustan Book Agency, 2007.
\bibitem{Schrodinger}
	E. Schrodinger, \emph{Uber die Unanwendbarkeit der Geometrie im Kleinen}, Naturwiss. 22: 518-520, 1934.
\bibitem{Heisenberg}
	W. Heisenberg, \emph{Uber die in der Theorie der Elementarteilchen Auftretende Universelle Lange}, Ann. Phys. 32: 20-33, 1938.
\bibitem{CDS}
	A. Connes, M. R. Douglas and A. Schwarz, \emph{Noncommutative Geometry and Matrix Theory: Compactification on Tori}, JHEP, no. 02, 003, 42 pages, 1998 [arXiv:hep-th/9711162].
\bibitem{Witten}
	E. Witten, \emph{Noncommutative Geometry and String Field Theory}, Nucl. Phys. B268: 253-294, 1986.
\bibitem{Veneziano}
	G. Veneziano, \emph{A String Nature Needs Just Two Constants}, Europhys. Lett. 2: 199-204, 1986.
\bibitem{GM}
	D. J. Gross and P. F. Mende, \emph{String Theory Beyond the Planck Scale}, Nucl. Phys. B303: 407-457, 1988.
\bibitem{ACV}
	D. Amati, M. Ciafaloni and G. Veneziano, \emph{Can Spacetime be Probed Below the String Size?}, Phys. Lett. B216: 41-47, 1989.
\bibitem{SW}
	N. Seiberg and E. Witten, \emph{String Theory and Noncommutative Geometry}, JHEP, no. 09, 032, 93 pages, 1999 [arXiv:hep-th/9908142].
\bibitem{Polchinski}
	J. Polchinski, \emph{Dirichlet Branes and Ramond-Ramond Charges}, Phys. Rev. Lett. 75: 4724, 1995 [arXiv:hep-th/9510017].
\bibitem{Witten2}
	E. Witten, \emph{Bound States of Strings and p-Branes}, Nucl. Phys. B460: 33, 1996 [arXiv:hep-th/9510135].
\bibitem{Sheikh-Jabbari}
	M. M. Sheikh-Jabbari, \emph{More on Mixed Boundary Conditions and D-branes Bound States}, Phys. Lett. B 425: 48-54, 1998 [arXiv:hep-th/9712199].
\bibitem{Snyder1}
	H. S. Snyder, \emph{Quantized Space-time}, Phys. Rev. 71: 38-41, 1947.
\bibitem{Snyder2}
	H. S. Snyder, \emph{The Electromagnetic Field in Quantized Space-time}, Phys. Rev. 72: 68-71, 1947.
\bibitem{Weyl}
	H. Weyl, \emph{The Theory of Groups and Quantum Mechanics}, Dover, New York, 1931.
\bibitem{Wigner}
	E. P. Wigner, \emph{On the Quantum Corrections for Thermodynamic Equilibrium}, Phys. Rev. 40: 749-759, 1932.
\bibitem{Groenewold}
	H. J. Groenewold, \emph{On the Principles of Elementary Quantum Mechanics}, Physica 12: 405-460, 1946.
\bibitem{Moyal}
	J. E. Moyal, \emph{Quantum Mechanics as a Statistical Theory}, Proc. Cambridge Phil. Soc. 45: 99-124, 1949.
\bibitem{BFFLS1}
	F. Bayen, M. Flato, C. Fronsdal, A. Lichnerowicz and D. Sternheimer, \emph{Deformation Theory and Quantization I. Deformation of Symplectic Structures}, Ann. Phys. NY, 111: 61-110, 1978.
\bibitem{BFFLS2}
	F. Bayen, M. Flato, C. Fronsdal, A. Lichnerowicz and D. Sternheimer, \emph{Deformation Theory and Quantization II. Physical Applications}, Ann. Phys. NY, 111: 111-151, 1978.
\bibitem{Filk}
	T. Filk, \emph{Divergencies in a Field Theory on Quantum Space}, Phys. Lett. B376: 53-58, 1996.
\bibitem{MVS}
	S. Minwalla, M. Van Raamsdonk and N. Seiberg, \emph{Noncommutative Perturbation Dynamics}, JHEP, no. 02, 020, 31 pages, 2000 [arXiv:hep-th/9912072].
\bibitem{Hayakawa1}
	M. Hayakawa, \emph{Perturbative Analysis on Infrared Aspects of Noncommutative QED on $R^4$}, Phys. Lett. B478: 394-400, 2000 [arXiv:hep-th/9912094].
\bibitem{Hayakawa2}
	M. Hayakawa, \emph{Perturbative Analysis on Infrared and Ultraviolet Aspects of Noncommutative QED on $R^4$}, Osaka 2000, High Energy Physics, vol. 2: 1455-1460 [arXiv:hep-th/9912167].
\bibitem{GKW}
	H. Grosse, T. Krajewski and R. Wulkenhaar, \emph{Renormalization of Noncommutative Yang-Mills theories: A Simple Example}, 2000 [arXiv:hep-th/0001182].
\bibitem{MST}
	A. Matusis, L. Susskind and N. Toumbas, \emph{The IR/UV Connection in the Non-commutative Gauge theories}, JHEP, no. 12, 002, 18 pages, 2000 [arXiv:hep-th/0002075].
\bibitem{GW1}
	H. Grosse and R. Wulkenhaar, \emph{Renormalization of $\phi^4$-Theory on Noncommutative $R^2$ in the Matrix Base}, JHEP, no. 12, 019, 26 pages, 2003 [arXiv:hep-th/0307017].
\bibitem{GW2}
	H. Grosse and R. Wulkenhaar, \emph{Renormalization of $\phi^4$-Theory on Noncommutative $R^4$ in the Matrix Base}, Commun. Math. Phys. 256: 305-374, 2005 [arXiv:hep-th/0401128].
\bibitem{GMRT}
	R. Gurau, J. Magnen, V. Rivasseau and A. Tanasa, \emph{A Translation-Invariant Renormalizable Noncommutative Scalar Model}, Comm. Math. Phys. 287:275-290, 2009 [arXiv:0802.0791 [math-ph]].
\bibitem{GLV}
	S. Galluccio, F. Lizzi and P. Vitale, \emph{Translation Invariance, Commutation Relations and Ultraviolet/infrared Mixing}, JHEP, no. 09, 054, 18 pages, 2009 [arXiv:0907.3640 [hep-th]].
\bibitem{Galluccio}
	S. Galluccio, \emph{Non-Commutative Field Theory, Translation Invariant Products and Ultraviolet/Infrared Mixing}, PhD thesis, 2010 [arXiv:1004.4655 [hep-th]].
\bibitem{Varshovi}
    A. A. Varshovi, \emph{Translation-Invariant Noncommutative Gauge Theories, Matrix Modeling and Noncommutative Geometry}, 2011 [arXiv:1101.3147 [hep-th]].
\bibitem{TV}
	A. Tanasa and P. Vitale, \emph{Curing the UV/IR Mixing for Field Theories with Translation-Invariant $\star$ Products}, Phys. Rev. D 81, 065008, 12 pages, 2010 [arXiv:0912.0200 [hep-th]].
\bibitem{Bayen}
	F. Bayen, \emph{In Group theoretical Methods in Physics}, ed. E. Beiglbouk et. al., Lect. Notes Phys. 94: 260-271, 1979.
\bibitem{Voros}
	A. Voros, \emph{Wentzler-Kramers-Brillouin method in the Bargmann representation}, Phys. Rev. A 40: 6814-6825, 1989.
\bibitem{BW1}
	M. Bordemann and S. Waldmann, \emph{A Fedosov star product of Wick type for Kahler Manifolds}, Lett. Math. Phys. 41, 243-253, 1997 [arXiv:q-alg/9605012].
\bibitem{BW2}
	M. Bordemann and S. Waldmann, \emph{Formal GNS Construction and States in Deformation Quantization}, Comm. Math. Phys. 195: 549-583, 1998 [arXiv:q-alg/9607019].
\bibitem{Rieffel}
	M. A. Rieffel, \emph{Deformation Quantizations for Actions of  $R^d$}, Mem. Am. Math. Soc. 106, no. 506, 93 pages, 1993.
\bibitem{Madore1}
	J. Madore, \emph{The Commutative Limit of a Matrix Geometry}, J. Math. Phys. 32: 332-335, 1991.
\bibitem{Madore2}
	J. Madore, \emph{The Fuzzy Sphere}, Class. Quantum Grav. 9: 69-87, 1992.
\bibitem{MR}
	S. Majid and H. Ruegg, \emph{Bicrossproduct Structure of $\kappa$-Poincare Group and Non-commutative Geometry}, Phys. Lett. B 334: 348-354, 1994 [arXiv:hep-th/9405107].
\bibitem{DJMTWW}
	M. Dimitrijevic, L. Jonke, L. Moller, E. Tsouchnika, J. Wess and M. Wohlgenannt, \emph{Deformed Field Theory on $\kappa$-Spacetime}, Eur. Phys. J. C 31: 129-138, 2003 [arXiv:hep-th/0307149].
\bibitem{RTF}
	N. Reshetikhin, L. Takhtadzhyan and L. Faddeev, \emph{Quantization of Lie Groups and Lie Algebras}, Leningrad Math. J. 1: 193-225, 1990.
\bibitem{Kassel}
	C. Kassel, \emph{Quantum Groups}, Graduated Texts in Mathematics 155, Springer-Velag, New York,1995.
\bibitem{Warner}
	F. W. Warner, \emph{Foundations of Differentiable Manifolds and Lie Groups}, Graduated Texts in Mathematics 94, Springer-Verlag, 1983.
\bibitem{Varshovi2}
    A. A. Varshovi, \emph{Groenewold-Moyal Product, $\alpha^*$-Cohomology, and Classification of Translation-Invariant Non-commutative Structures}, 2012 [arXiv:1210.0695 [math-ph]].
\bibitem{HW}
	G. H. Hardy and E. M. Wright, \emph{An Introduction to the Theory of Numbers}, Oxford, 1954.
\end{thebibliography}
\end{document}